\documentclass[]{mn2e}
\usepackage{times,epsfig,lscape,rotating}

\title[Diffuse calibration of  {\it IRAC} through study of HII regions]
{Absolute diffuse calibration of  {\it IRAC} through mid-infrared and radio study of H{\sc ii} regions}
\author [Martin Cohen et al.]
{Martin Cohen$^{1}$\footnote{email: mcohen@astro.berkeley.edu}, Anne J. Green$^{2}$, Marilyn R. Meade$^{3}$, Brian Babler$^{3}$
\newauthor
R\'emy Indebetouw$^{4}$, Barbara A. Whitney$^{5}$, Christer Watson$^{6}$, Mark Wolfire$^{7}$
\newauthor
Mike J. Wolff$^{5}$, John S. Mathis$^{3}$, and Edward B. Churchwell$^{3}$\\
$^{1}$Radio Astronomy Laboratory, University of California, Berkeley, CA 94720, USA\\ 
$^{2}$School of Physics, University of Sydney, NSW 2006, Australia\\
$^{3}$Dept. of Astronomy, University of Wisconsin, Madison, WI 53706, USA\\
$^{4}$Astronomy Dept. University of Virginia, Charlottesville, VA 22904, USA\\
$^{5}$Space Science Institute, Boulder, CO 80303, USA\\
$^{6}$Dept. of Physics, Manchester College, North Manchester, IN 46962, USA\\
$^{7}$Dept. of Astronomy, University of Maryland, College Park, MD 20742, USA\\ }

\date{ Accepted . Received ; in original form        }

\begin{document}

\maketitle
\clearpage
\begin{abstract}
We investigate the diffuse absolute calibration of the InfraRed Array Camera on the {\it Spitzer} 
Space Telescope at 8.0\,$\mu$m using a sample of 43 H{\sc ii} regions with a wide range of
morphologies near {\it l}=312$^\circ$.  For each region we carefully measure sky-subtracted,
point-source-subtracted, areally-integrated {\it IRAC} 8.0-$\mu$m fluxes and compare these with 
Midcourse Space eXperiment (MSX) 8.3-$\mu$m images at two different spatial resolutions, and with radio 
continuum maps.  We determine an accurate median ratio of {\it IRAC}\,8.0-$\mu$m/MSX\,8.3-$\mu$m fluxes, 
of 1.55$\pm$0.15.  From robust spectral energy distributions of these regions we
conclude that the present 8.0-$\mu$m diffuse calibration of the SST is 36\,percent too high compared 
with the MSX validated calibration, perhaps due to scattered light inside the camera.  This is 
an independent confirmation of the result derived for the diffuse calibration of 
IRAC by the {\it Spitzer} Science Center (SSC).

From regression analyses we find that 843-MHz radio fluxes of H{\sc ii} regions 
and mid-infrared (MIR) fluxes are linearly related for MSX at 8.3\,$\mu$m and {\it Spitzer} 
at 8.0\,$\mu$m, confirming the earlier MSX result by Cohen \& Green.  The median ratio of 
MIR/843-MHz diffuse continuum fluxes is 600$\times$ smaller in nonthermal than 
thermal regions, making it a sharp discriminant.  The ratios are largely independent of 
morphology up to a size of $\sim24^{\prime}$. We provide homogeneous 
radio and MIR morphologies for all sources.  MIR morphology is not 
uniquely related to radio structure.  Compact regions may have MIR filaments and/or diffuse 
haloes, perhaps infrared counterparts to weakly ionized radio haloes found around compact H{\sc ii} regions.  
We offer two {\it IRAC} colour-colour plots as quantitative diagnostics of diffuse H{\sc ii} regions.  
\end{abstract}


\section{INTRODUCTION}
Cohen \& Green (2001: hereafter CG) made a detailed comparison of MIR and radio continuum 
imaging of H{\sc ii} regions in an 8~deg$^2$ Galactic field centred near $l = 312^\circ$.
Using 8.3-$\mu$m images from the survey of the Galactic Plane by the Midcourse Space eXperiment 
(MSX: Price et al. 2001) and 843-MHz continuum maps from the Molonglo Observatory Synthesis Telescope 
(MOST: Green et al. (1999)), CG developed a ratio of spatially integrated MIR to radio
flux densities that discriminates between objects with thermal and nonthermal radio emission.
This field was first explored at 843~MHz by Whiteoak, Cram \& Large (1994: hereafter WCL) who
characterized the emission of sources on all scales as thermal or nonthermal.  The inventory of 
this region is diverse.  WCL and CG list more than fifty thermal sources, ranging from compact 
H{\sc ii} regions to filaments extending over 1$^\circ$ in length.  Only about one third of the
objects were detected in the recombination line survey of Caswell \& Haynes (1987).  Very few 
lie at tangent points so that no distance estimates, or only ambiguous ones, exist for most
of the sources.

The {\it Spitzer} Space Telescope (Werner et al. 2004: hereafter SST) GLIMPSE (Galactic Legacy 
Infrared Mid-Plane Survey Extraordinaire) program (Benjamin et al. 2003; Churchwell et al. 2004) 
has surveyed 220~deg$^2$ of the plane in all four bands of the InfraRed Array Camera ({\it IRAC}: 
Fazio et al. 2004).  GLIMPSE images of the 
plane are roughly 1000 times deeper, and have an areal resolution of order 100 times 
smaller, than those of MSX.  The SST offers the prospect of exploring faint structure 
unseen by MSX.  

This study has several different objectives. All are achievable by combining radio and MIR continuum 
surveys of hundreds to thousands of square degrees of the Galactic Plane having the best resolutions 
currently available for such panoramic coverage.  {\it Spitzer} provides GLIMPSE images of 
limited portions of the plane at high spatial resolution.
First, we quantitatively investigate the diffuse calibration of {\it IRAC} using 
the photodissociation regions (PDRs) that envelop thermal radio regions.  Our purpose is to 
attempt to link {\it IRAC}'s 8.0-$\mu$m diffuse calibration to that achieved by MSX and 
validated absolutely to $\sim$1\,percent by Price et al. (2004).  Absolute calibration was a vital
activity on MSX, which had both point source and diffuse calibrations.  Spatially integrating 
images of point sources and comparing those results with the MSX point source catalogs shows
these two calibrations are consistent.  Therefore, MSX serves as a benchmark.
Secondly, we confirm CG's median ratio of MSX\,8.3-$\mu$m to 843-MHz flux densities, $F_{8.3}/S_{843}$,
for thermal emission, using a broader assessment of sky background than CG used, and investigate whether this 
discriminant varies with the density, angular scale, and morphology of thermal radio sources.
Thirdly, we extend the MSX/radio discriminant between thermal and nonthermal radio emission
to the equivalent {\it IRAC}\,8.0-$\mu$m ratio, after allowance for the different radiance 
contributions of polycyclic aromatic hydrocarbon (PAH) emission in these two different filters.    
Fourthly, we examine {\it IRAC} images of compact H{\sc ii} regions looking for evidence of MIR 
proxies for the weakly ionized radio haloes created by the leakage of photons from the dense cores. 
This phenomenon was proposed to explain the existence of extended radio continuum emission associated 
with ultracompact H{\sc ii} regions (e.g. Kurtz et al. 1999).  Fifth, we compare the 
MIR spectral energy distributions (SEDs) 
of H{\sc ii} regions in the field.  Finally, we note that there are other radio 
structures in the field that have faint MIR counterparts that were not recognized by 
MSX because of its poorer resolution.  One example is the optically invisible planetary 
nebula PNG313.3+00.3, well-detected by GLIMPSE (Cohen et al. 2005).  We draw attention to another
particularly interesting object and demonstrate that it too is a thermal radio emitter.

\S2 introduces the theme of the absolute calibration of {\it IRAC}, explaining the origin
of the differences between point source and diffuse calibration.
In \S3 we detail the 43 thermal objects studied by CG (eliminating all sources known to 
be galaxies or to have nonthermal radio spectra).  In \S4 we describe our methods for
assessing spatially integrated fluxes in MIR and radio images, and compare 
MSX\,8.3-$\mu$m integrated fluxes with the equivalent {\it IRAC}\,8.0-$\mu$m values as a function 
of H{\sc ii} region morphology.  \S5 uses these SST/MSX ratios to probe the
diffuse calibration of {\it IRAC} based on MIR spectra of our target H{\sc ii} regions.
\S6 discusses the SEDs among the population of H{\sc ii} regions in this field, expressed through their 
IRAC colours. \S7 details an improved set of ratios of MSX\,8.3-$\mu$m to MOST 843-MHz flux densities
and provides the corresponding ratios constructed using the current calibration for GLIMPSE 
8.0-$\mu$m image products.  Statistical regression analyses are also given.   
\S8 illustrates {\it IRAC} imagery of some of the regions studied, highlighting 
similarities and differences between MIR and radio morphologies.  \S9 presents 
a curious thermal object with unique characteristics. \S10 provides a MIR/radio ratio
for five galaxies and one supernova remnant (SNR) lying within this field, to illustrate
the difficulty of measuring the MIR emission of SNRs even with the SST, but 
also to provide a quantitative estimate of the MIR/radio ratio for nonthermal emitters.
\S11 offers our conclusions.

\section{THE ABSOLUTE CALIBRATION OF {\it IRAC}}
The basis of {\it IRAC}'s absolute calibration is identical to the network of fiducial standards
built by Cohen et al. (1999), founded on Sirius.  This calibration has been validated
absolutely by Price et al. (2004) to 1.0\,percent.  To support calibration with the high sensitivity
of {\it IRAC} required the development of new techniques and the establishment of faint optical-to-IR
calibrators (Cohen et al. 2003).  {\it IRAC} standards are either K0-M0III or A0-5V stars.  The
former are stellar supertemplates (UV-optical versions of the original cool giant templates
of Cohen et al. (1999)).  The latter are represented by Kurucz photospheric spectra.  These
stars underpin the calibration of {\it IRAC} and provide cross-calibration pairwise between the three
instruments of the SST.  The primary {\it IRAC} suite consists of 11 stars near the North Ecliptic
Pole, but a secondary network close to the Ecliptic Plane offers efficient checks of calibration
every 12\,hr after {\it IRAC} data are downlinked.  Reach et al. (2005) express their preference for
the A-stars and it is these that provide the fundamental parameters that are embedded in the
header of each item of Basic Calibrated Data (BCD) from the SST.  Excellent and detailed 
summaries of {\it IRAC}'s calibration are given by Fazio et al. (2004), Reach et al. (2005), and in
the {\it IRAC} Data Handbook (version 3.0, available from the SSC.)  

Calibration of point sources is achieved by performing aperture photometry on BCD images,
in a 10-pixel (12.2$^{\prime\prime}$) radius aperture, and comparing with observations
of the {\it IRAC} standards using the same process, aperture size, and sky annulus.  Treating BCD images in 
this fashion results in consistent, well-calibrated, point source photometry that traces to MSX
(Cohen et al. 2003).  It had been assumed
before the launch of the SST that the extended emission calibration for {\it IRAC} could be derived 
from the point source calibration.  Several tasks were planned to check the accuracy of the 
extended source calibration, but only after launch was it determined that the 
responsivity to point source emission was lower than expected, and that differences existed
between point source and extended source calibration, consistent with internal scattering 
of the light in the {\it IRAC} arrays.  As a consequence, extended emission will not be correctly
calibrated and must be corrected to the ``infinite aperture" limit by multiplying by 
wavelength-dependent ``effective aperture correction" factors.  The factor advocated for {\it IRAC} 
extended sources at 8.0\,$\mu$m is 0.737 (IRAC Data Handbook, Table 5.7; Reach et al. 2005). 
The recommended factors for {\it IRAC}'s three shorter wavelength bands are
0.944, 0.937, 0.772 at 3.6, 4.5, 5.8\,$\mu$m, respectively.  In this paper we investigate
the correction factor at 8.0\,$\mu$m by an independent approach.

\section{H{\sc ii} REGION INVENTORY OF THE FIELD}
Table~1 lists the thermal radio objects in the WCL study of the field.  Thirteen sources 
listed by CG  were excluded from our primary analysis, eleven of them nonthermal emission
regions: five galaxies, four known SNR, a further two nonthermal radio sources (one of 
them mixed inextricably with thermal emission), one source of unknown nature that is surely 
dominated by thermal dust emission (G313.28$-$0.33), and the source G311.31+0.60 
(which is a misnomer by WCL for G311.42+0.60 and appears in CG as a duplicate of this 
latter source's entry.  G310.80$-$0.38 cannot be isolated unambiguously in the complex 
{\it Spitzer} images).  We have also inserted one new source, G311.70+0.31, of particular 
interest because of its unique morphology (\S9).
We initially adopted the source character (e.g. classical, compact, filamentary, etc.) from WCL.
However, these attributes were drawn from heterogeneous literature so we decided to re-examine
the Molonglo data for every source and have assigned our own homogeneous radio types to the sample. 
These are indicated in Table~1 (col.(3)), as are WCL's radio dimensions in arcmin (col.(2):
Galactic longitude by latitude).  Columns (5), (6), and (7) respectively present the observed ratios of 
the {\it IRAC}\,8.0-$\mu$m to MSX\,8.3-$\mu$m flux densities, and of the MSX\,8.3-$\mu$m and 
{\it IRAC}\,8.0-$\mu$m flux densities to those at 843~MHz.  Uncertainties are given for each ratio 
(1$\sigma$) based upon the root-sum-square (RSS) of the fractional errors in the pairs of flux 
densities.  Column (8) gives the areally integrated 8.0-$\mu$m SST fluxes in mJy. Column (9) gives 
figure numbers for objects represented by images.

SST offers MIR images with
far higher spatial resolution than those of currently available panoramic radio surveys.  Therefore, we
have also developed a MIR morphology based on the {\it IRAC} images (some of which are presented in \S8),
which appears in col.(4) of Table~1.  Note that G311.64+0.97, a bubble some 35$\times$30 arcmin, extends
beyond the northern latitude coverage of GLIMPSE.  For the calculations in this paper we have used 
the brightest portion that can be accommodated by the GLIMPSE latitude limit, namely the 
6$\times$14 arcmin eastern filament centred at (311.88$^\circ$,1.01$^\circ$).

For convenience we have ordered sources in Table~1 by longitude.  To render the table more readable 
we have selected a small number of morphological descriptors that are coded  by
their leading letters.  In the MOST radio continuum these are: C - compact; S - shell (curved
morphologies); B - blob (lacking any specific structure but larger than a compact or
unresolved source; and F - filamentary.  The {\it IRAC} descriptors are: UC - ultracompact; 
C - compact; S - shell; D - diffuse; F - filamentary.  However, it  was necessary to 
add a second (lower-case) letter to signify secondary characteristics recognizable in the 
detailed imagery.  These extra qualifiers are: c - compact; m - multiple (when multiple
compact cores occur within a source; s - shell; f - filaments present; h - halo;
i - objects that are apparently in close enough physical proximity to be interacting.  

The scheme we have adopted, based on GLIMPSE, must clearly represent   
those characteristics that are detected in this particular survey. A
deeper survey might well be expected to discover fainter features
unknown from GLIMPSE.  A future change in pipeline, or in our understanding
of diffuse calibration, could readily alter underlying background levels
and local brightnesses, thereby affecting areal integrations of specific 
parts of an H{\sc ii} region.
Similarly, longer exposures could reveal fainter structures that, at 
GLIMPSE's depth, appear separate and distinct but, with greater depth in
imaging, are found to be connected by filaments or to be merely small
portions of widespread faint haloes.  Therefore, we rejected the ratio of 
components' integrated fluxes as the determining factor for morphological
types. Instead we have assigned types by ranking the highest 
surface brightnesses found in continuous topological elements in residual
images, typically at 8.0\,$\mu$m.  

For example, Fig.~\ref{31148} (in \S8) shows compact cores, a large, faint, partial 
shell, many separate, small, internal shells, filaments, and even diffuse emission.
The highest surface brightness pixels occur in the very bright central core.
The second highest levels are found in the small, complete rings north of the
bright core, and the third in the long, curving filament that emerges to the
north-west from the core, looping around to the north and almost reaching
the brighter of the two small rings inside the large, faint oval shell.  Hence,
``Cs" is our designation.
Tertiary characteristics are not given for simplicity so the secondary code
describes whichever aspect is brightest after the primary.  Had we used a third
descriptor then G311.48+0.37 would have been ``Csf". Parentheses around a primary code imply
that it is not definite that the MIR source is the counterpart of the radio object.

\begin{table*}   
\begin{center}
{\bf Table~1}. The sample of H{\sc ii} regions with radio and MIR morphologies, integrated MIR and radio flux ratios, 
and 8.0-$\mu$m fluxes
\vspace{3mm}
\begin{tabular}{lcccllrrc}
Source& Size& Radio& {\it IRAC}& {\it IRAC}/&  MSX/& {\it IRAC}/& {\it IRAC}& Fig.\\
   &(arcmin)& morph.& morph.& MSX& MOST& MOST& mJy& \\
\hline
G310.63--0.43&  $<$1&          C&   Cf&    1.64$\pm$0.20&  23$\pm$1&    37$\pm$5&       17600& \\
G310.68--0.44&  $<$1&          C&   Cf&    1.49$\pm$0.20&  23$\pm$1&    34$\pm$4&       13800& \\
G310.69--0.31&  $<$1&          C&   UC&    1.16$\pm$0.15&  5.0$\pm$0.2& 6$\pm$1&         5100& \\
G310.80--0.38&  7$\times$6&    S&   Dc& ...&            8$\pm$1& ...&                     ...& \\
G310.89+0.01&   $<$1&          C&   Cs&    0.93$\pm$0.12&  37$\pm$2&    34$\pm$4&        7200& 18\\
G310.90--0.38&  $<$1&          C&   Ch&      1.10$\pm$0.15&  11$\pm$1&  12$\pm$1&        9600& \\
G310.99+0.42&   5$\times$7&    S&   S& 2.02$\pm$0.29&  120$\pm$10&  250$\pm$30&        363000& 23\\
G311.12--0.28&  7$\times$8&    S&   Dc&    0.98$\pm$0.13&  15$\pm$1&  15$\pm$2&         50000& \\
G311.20+0.75&   6$\times$15&   S&   Cf& 3.63$\pm$0.63&  46$\pm$6&    170$\pm$20&       325000& \\
G311.29--0.02&  15$\times$14&  F&   Dc&  6.27$\pm$0.78&  40$\pm$4&    250$\pm$40&      831000& \\
G311.30+0.90&   8$\times$4&    S&   D&   1.14$\pm$0.15&  72$\pm$2&    82$\pm$10&        64000& \\
G311.37+0.79&   8$\times$24&   F&   Dc& 1.73$\pm$0.27&  87$\pm$9&    150$\pm$20&       298000& \\
G311.42+0.60&   $<$1&          C&   Cs&    1.83$\pm$0.23&  74$\pm$5&    140$\pm$20&     25000& 16\\
G311.48+0.37&   5$\times$8&    S&   Cs&    1.19$\pm$0.16&  24$\pm$1&    29$\pm$4&      129000& 20\\
G311.50--0.48&  8$\times$8&    S&   D&  2.16$\pm$0.29&  15$\pm$0.5&  32$\pm$4&         111000& 24\\
G311.53--0.13&  2$\times$9&    F&   Df&  1.93$\pm$0.26&  20$\pm$1&    38$\pm$5&         21000& \\
G311.59--0.59&  $<$1&          C&   Cf&  1.61$\pm$0.20&  17$\pm$1&  28$\pm$3&            7600& 17\\
G311.62--0.60&  11$\times$12&  S&   Dc& 1.31$\pm$0.18&  66$\pm$7&    86$\pm$10&        189000& \\
G311.63+0.29&   $<$1&          C&   Cm&    2.63$\pm$0.32&  13$\pm$1&  34$\pm$4&         32800& \\
G311.64+0.97&   35$\times$30$^a$&  F&   F& 1.70$\pm$0.23&  48$\pm$5&  82$\pm$10&         43600& \\
G311.70+0.31&   3$\times$2&    F&   C(\S9)&   1.71$\pm$0.32&  27$\pm$5&    46$\pm$7&          4800& \\
G311.87--0.24&  $<$1&          C&   Ch&     1.61$\pm$0.21&  18$\pm$1&  29$\pm$3&         9000& \\
G311.90+0.08&   6$\times$6&    S&   Cf&     2.30$\pm$0.31&  26$\pm$2&    59$\pm$7&     340000& \\
G311.93+0.21&   6$\times$5&    S&   Cf&    1.64$\pm$0.22&  25$\pm$1&    42$\pm$5&      274000& \\
G311.97--0.05&   6$\times$4&   B&   Dc&   2.30$\pm$0.46&  33$\pm$6&    75$\pm$11&       65700& \\
G312.11+0.31&   $<$1&          C&   Cs&    1.47$\pm$0.20&  28$\pm$1&  41$\pm$5&         42400& \\
G312.36--0.04&  $<$1&          C&   Df& 1.13$\pm$0.19&  24$\pm$3&    28$\pm$3&          13200& \\
G312.45+0.08&   60$\times$90&  F&   F&  1.46$\pm$0.35&  800$\pm$200&   1200$\pm$230&  9460000& 26\\
G312.50+0.32&   5$\times$10&   F&   D& 1.56$\pm$0.24&  53$\pm$6&   83$\pm$10&           68500& \\
G312.58+0.40&   8$\times$7&    B&   Cm&  1.85$\pm$0.24&   37$\pm$2&    69$\pm$9&        57000& \\
G312.59+0.22&   3$\times$4&    B&   Df& 2.34$\pm$0.40&  29$\pm$4&    68$\pm$10&         45800& \\
G312.60+0.05&   $<$1&          C&   Ch&   2.46$\pm$0.31&  48$\pm$2&    120$\pm$15&      15700& 19\\
G312.67--0.12&  3$\times$1&    B&   S&     1.46$\pm$0.24&  57$\pm$8&    83$\pm$12&      12300& 22\\
G312.68+0.04&   3$\times$4&    Ci&  Si&    2.39$\pm$0.35&  31$\pm$3&    75$\pm$10&      38300& 21\\
G312.71+0.02&   3$\times$3&    Ci&  Si&  1.05$\pm$0.15&  33$\pm$4&    35$\pm$5&         18000& 18\\
G312.72--0.14&  $<$1&          C&   (S)&    0.98$\pm$0.13&  83$\pm$2&    82$\pm$10&     15900& \\
G312.77+0.06&   $<$1&          C&   Ch&   1.94$\pm$0.25&  250$\pm$20&  500$\pm$60&      28600& 15\\
G312.93--0.09W& 8$\times$9&    S&   D&    1.75$\pm$0.23&  59$\pm$4&    100$\pm$15&      88000& \\
G312.95--0.44&  5$\times$4&    S&   Si&    1.06$\pm$0.18&  40$\pm$5&    43$\pm$5&       60900& \\
G313.07+0.32&   5$\times$4&    S&   Df&    1.24$\pm$0.26&  64$\pm$15&    79$\pm$15&     30000& 25\\
G313.32+0.05&   8$\times$3&    F&   D&   0.98$\pm$0.13&  30$\pm$2&    21$\pm$3&         60600& \\
G313.37+0.02&   9$\times$3&    S&   Dc&   1.27$\pm$0.17& 74$\pm$1& 94$\pm$11&           37700& \\
G313.46+0.19&   $<$1&          C&   Ch&    1.79$\pm$0.24&  16$\pm$1&  28$\pm$3&         22000& \\
\end{tabular}
\end{center}
Footnote: $^a$ We used the maximum bright area of 6$\times$14\,arcmin of this region that was covered by GLIMPSE
\end{table*}

\section{COMPARISON OF SST AND MSX INTEGRATED FLUXES}
CG used matched-resolution MSX and radio images to derive their MIR-to-radio ratios.  
The Galactic Plane is particularly rich in both local and more distant structures
which emit in the PAH bands.  Accurate  photometry of extended sources in the presence 
of this clutter requires special care.  This affects MSX\,8.3-$\mu$m measurements 
and potentially three {\it IRAC} bands (3.6, 5.8, and 8.0\,$\mu$m) containing PAH fluorescent
features at 3.3, 6.2, 7.7, and 8.7\,$\mu$m.  The coarseness of the low-resolution 
MSX images could have caused us to discriminate poorly between faint MIR emission and
sky background.  For example, we might have sampled inappropriately bright regions to 
assess the sky emission, 
particularly if we now wish to seek faint outer haloes.  Therefore, we have twice 
remeasured all regions in the images with 46$^{\prime\prime}$ resolution 
(36$^{\prime\prime}$ pixels), paying particular attention to the choice of local sky 
background and using more than a single local area to assess the variations in sky
background whenever possible.  We have repeated this same procedure on the
20$^{\prime\prime}$ resolution (6$^{\prime\prime}$ pixel) ``high-resolution" 
MSX images, better to assess the detectable 
extent of the sources, and with the same considerations of the sky emission to 
be subtracted.  In this paper, we use the average of these two 
sets of measurements to provide our best estimates of the  8.3-$\mu$m 
flux densities of these extended regions.  We have quantified the likely 
uncertainties in these integrals using the dispersions of the several sets of
sky-subtracted measurements for each source.  The average fractional uncertainty 
in these 8.3-$\mu$m areal integrations was found to be 6\,percent.

For the {\it IRAC} fluxes in all four bands we have likewise sought to include all IR emission 
apparently associated with the radio sources in our spatial integrals, and to 
identify background that is well outside the radio sources.  Unlike the MSX images, 
we have worked from GLIMPSE ``residual images".  These are 
1.1$^{\circ}$$\times0.8^\circ$ images with 0.6$^{\prime\prime}$ pixels 
from which all GLIMPSE point sources have been removed.  The residual images 
are ideal for enhancing the recognition of diffuse nebulosity in regions of 
high point source density.  They enable far more reliable photometry of 
such emission than can be obtained in the presence of the many point sources 
in the field.  The residual images reflect the subtraction by our adaptation of 
DAOPHOT for GLIMPSE of all sources detected down to 2$\sigma$, deeper than 
our publically released Catalog and Archive point source lists that extend down 
only to 10 and 5 sigma, respectively.  Thus there are faint 2$-$3-$\sigma$ points 
that are subtracted from the residual images that are not listed anywhere
in our lists.  Residual images still contain sources not extracted by DAOPHOT, such 
as saturated sources and sources that peak beyond the non-linearity limit for each 
band.  These objects were individually integrated and their flux densities 
subtracted from the integrals over the relevant regions.  (Fig.~\ref{dragon} illustrates
a residual image.)  PSF-subtracted residual artefacts consist of both negative
and positive portions, often with a near-zero net integral.  These usually
have very little effect on integrations of diffuse emission but they can make 
contributions of order 1$-$10\,percent to small areas of integrated sky background.
Therefore, we removed all such artefacts that lay within regions we took to assess 
the sky, and likewise subtracted the brightest recognizable artefacts lying within 
the boundaries of H{\sc ii} regions.

\subsection{A statistical approach to removing MSX point sources}
To compare the truly diffuse component of emission for an H{\sc ii} region measured from 
MSX imagery with that assessed from the corresponding {\it IRAC} residual image one must
subtract the contribution made to the total MSX brightness by point sources.  This is 
required to avoid biasing the direct ratio of {\it IRAC} and MSX brightnesses by the inclusion
of faint point sources in the MSX images.  We emphasize that, were this the case, the ratio
obtained would be {\it lower} than the true value because the MSX brightness would have been
overestimated.  Ideally one should remove the identical point sources from MSX images as 
from the {\it IRAC} images but these are roughly 1000$\times$ deeper than MSX so the vast 
bulk of GLIMPSE point sources were completely undetected, and undetectable, by MSX.  

The majority of GLIMPSE sources are normal stars that appear faintest to {\it IRAC} in the
8.0-$\mu$m band, and were undetected by MSX.  Any MSX point sources in the H{\sc ii} regions 
that were discernible to the eye (i.e. at least 3$\sigma$) were integrated separately and 
their fluxes removed from that of the region.  The median number of point sources subtracted 
per region was four.  For a region with up to four point sources this approach was the 
simplest option and it was applied to more than half of our sample of H{\sc ii} regions.  
For the seventeen regions having five or more point sources we queried the latest MSX Point
Source Catalog (ver.2.3) using the ``Gator" search tool at the Infrared Processing and 
Analysis Center (IPAC) and summed the flux densities of the resulting table of sources
with detections within the specified area.  

In future investigations we would like to be able to remove the influences of point 
sources of MSX on wide area (tens to hundreds of square degrees) integrated fluxes as 
accurately as possible, including objects with signal-to-noise values of less than 
$3-5\sigma$.  Such weak objects cannot be drawn
from the MSX Reject Source Catalog due to the heterogeneity of this Catalog (that
can compromise the validity of the sources themselves) and to the effects of Malmquist 
bias (that overestimate the flux densities for valid weak sources).  Therefore, we have 
used the present paper as a formal test of 
a statistical method that relies on the use of the ``SKY" model originally presented by 
Wainscoat et al. (1992) but with considerably expanded capabilities.  Not only does the 
model reliably predict the LogN-LogS distribution of bright and faint point sources in 
arbitrary directions, in any user-supplied system band from the far-ultraviolet to the MIR,
it is also capable of summing up the smeared surface brightness due to unresolved and  
unrecognized point sources in any given area.  This ``Total Surface Brightness" mode was 
first described by Cohen (1993a), was applied to COBE/DIRBE diffuse sky maps (Cohen 1993b),
and has undergone testing since then in a variety of different spectral regimes (e.g. 
Cohen 2000).  Only one region in our sample contained so many MSX 8.3-um point 
sources that one could not remove them manually and one might not want to do so even by 
listing all sources in the MSX PSC2.3 and the Reject Source Catalog because of the likely
impact of Malmquist bias on the resulting summation of flux.  This area is G312.45+0.08, 
a 1.5$^\circ\times$1$^\circ$ area containing the largest diffuse structures in this field
and straddling the Galactic plane.  It is, therefore, well-chosen in terms of sampling the 
highest surface densities of point sources as well as containing much clutter in the form 
of bright nebulous filaments.
The newest version of the SKY model successfully predicts GLIMPSE  {\it IRAC} source 
counts by including the warp of the disk and was used for the calculations below.

The brightest and faintest stars in this field have magnitudes at 8.3\,$\mu$m of $-$0.06 
and 7.05, respectively (62\,Jy and 88\,mJy).  The statistical assessment of star counts in 
this 1.5 deg$^2$ area came from the computation that the total brightness of all point sources 
within these limits in the MSX 8.3-$\mu$m band amounts to 
5.44$\pm$0.22$\times$10$^{-11}$\,W m$^{-2}$deg$^{-2}$. Dividing by the MSX bandwidth 
and scaling by 1.5 for the total area observed yields 582$\pm$23\,Jy.  We have tested
these SKY predictions by directly finding all 8.3-$\mu$m point sources in the MSX
PSC2.3 that lie within the rectangle defined by G312.45+0.08, by a corresponding J2000
Equatorial polygon search using the {\it IPAC} Gator tool.  Eliminating all
sources that were undetected (i.e. have quality flag of 0 or 1) results in a total of
693 point sources whose accumulated flux density is 542$\pm$24\,Jy, statistically 
the same ($\sim1\sigma$) 
as SKY predicts.  We conclude that SKY is a viable tool for assessing total surface
brightness due to point sources in any region and we have utilized its capability
in this paper to ensure that effectively the same set of point sources has been
removed from the MSX images as from the {\it IRAC} residual images.  

SKY predicts more flux than that observed which is as expected because the MSX Point 
Source Catalog may not be complete to 7.05 magnitude because of the clutter of the 
interstellar medium (ISM) due to PAH emission.  Therefore, we also need a SKY 
calculation for MSX that mimics the DAOPHOT extraction process that generates
the {\it IRAC} residual images, by removing all point sources to the 2$\sigma$
level, corresponding to 8.0 magnitude.  That calculation yields 625$\pm$26\,Jy for the
MSX point sources that would have been subtracted to the same brightness level as in 
our {\it IRAC} residual images at 8.0\,$\mu$m.  The observed sum of diffuse and point 
sources in the box integrated is 7100\,Jy.  Therefore, point sources contribute only 
9\,percent of the integrated emission and the diffuse component is 6480\,Jy in this 
very large area.  For a more typical diffuse region in
our sample, with an area of $\sim$90\,arcmin$^2$, the point source contribution subtracted
from the total integrated light at 8.3\,$\mu$m is $\sim$6.5\,Jy, or 4\,percent.  Thirteen
of the fifteen compact and ultracompact sources have three or fewer recognizable MSX point
sources to magnitude 7.0, and their contribution is between 0.3 and 2\,percent of the dominating
emission of the ionized zone itself.  For the line-of-sight to the centre of each H{\sc ii}
region we ran the SKY model to compute the additional contribution made by point sources,
fainter than the limit to which we could recognize sources in the MSX image, and extending 
down to magnitude 8.0 at 8.3\,$\mu$ to match the sources already removed from the 
coresponding {\it IRAC} residual image.  The difference between assessing the MSX point
source light extrapolated by SKY to magnitude 8.0 and removing identified sources
to magnitude 7.0 is only about 7\,percent of the contribution of point sources
identified to magnitude 7.0.  Therefore, except for extremely large diffuse areas such
as G312.45+0.08, in individual H{\sc ii} regions the removal of recognizable MSX point
sources accounts for about $1-4$\,percent of the total measured diffuse emission.  
The model extrapolations, to match the depth of removal of point sources in the 
{\it IRAC} residual maps, imply the subtraction of only about an extra 1\,percent or 
less of the diffuse emission, well below the errors in the MIR integrated photometry
of the same regions.

We now summarize the corrections applied to the total integrated emission of regions for the
contribution of point sources.  The {\it IRAC} images from which we start the analysis
are already point-source subtracted to a depth of 8.0 magnitude at 8.0\,$\mu$m.  We
further remove artefacts by hand that correspond to slight mismatches between our PSFs and 
real point sources.  To subtract as closely as possible the same sources from MSX images
as were removed in constructing the {\it IRAC} residual images ,we used a combination of
techniques as follows, according to the character and apparent size of the H{\sc ii} region.
For compact and ultracompact objects we measured fluxes of all recognizable point sources
if fewer than five were seen in the image, and subtracted their total flux from the measured
diffuse emission.  For regions containing five or more point sources we extracted those
using Gator and summed their fluxes.  We extended every H{\sc ii} region below the local 
8.3-$\mu$m magnitude limit for unambiguous identification in our MSX images by using the 
SKY-predicted total of surface brightness due to faint point sources in that direction.
This contribution was subtracted from the photometry for each target region.  We believe that
this combination of techniques applied to the MSX images to account for point source removal is 
the best that can currently be achieved to match the depth of the IRAC residual images.  
If any starlight has been missed by the removal of actual sources and our extrapolation 
using SKY, it should be  negligible compared with the diffuse flux attributable to any given
H{\sc ii} region.

No statistical model of the sky, however physically realistic (as is SKY), can predict
the occurrence of individual young star clusters either along the line-of-sight or within
H{\sc ii} regions.  Indeed, objects in the bright cores of HII regions are observed on top
of bright nebular emission and will not appear pointlike.  Thus DAOPHOT will not 
identify them as point sources nor will they be removed from {\it IRAC} residual images.  
One might be concerned that such clusters would be treated differently by our handling
of point source removal from MSX and {\it IRAC} images.  However, our field contains no giant 
H{\sc ii} regions nor complexes with multiple star-forming centers, commonly the sites of 
embedded clusters.  None of our bright-cored regions, nor the more diffuse, open structured
regions reveals any projected or apparent internal clusters to {\it IRAC}.  This is largely 
because of the use of this field to demonstrate the capabilities of the MOST survey by WCL.
It is also a consequence of CG's choice to carry out their MIR-radio study, intended to   
focus on the ISM, and on interactions between the medium and individually dominant stars,
in this same field.  
Our analyses of MSX images, therefore, suffer from no bias with respect to those of IRAC that 
is caused by unresolved star clusters.

There are no low-resolution GLIMPSE 
images like the ones we used for MSX that offer quasi-independent photometry.  Based on 
a series of tests carried out on diverse H{\sc ii} regions of the reproducibility of our measured
sky-subtracted fluxes we have assigned

\onecolumn
\begin{figure}
\vspace{9.5cm}
\includegraphics{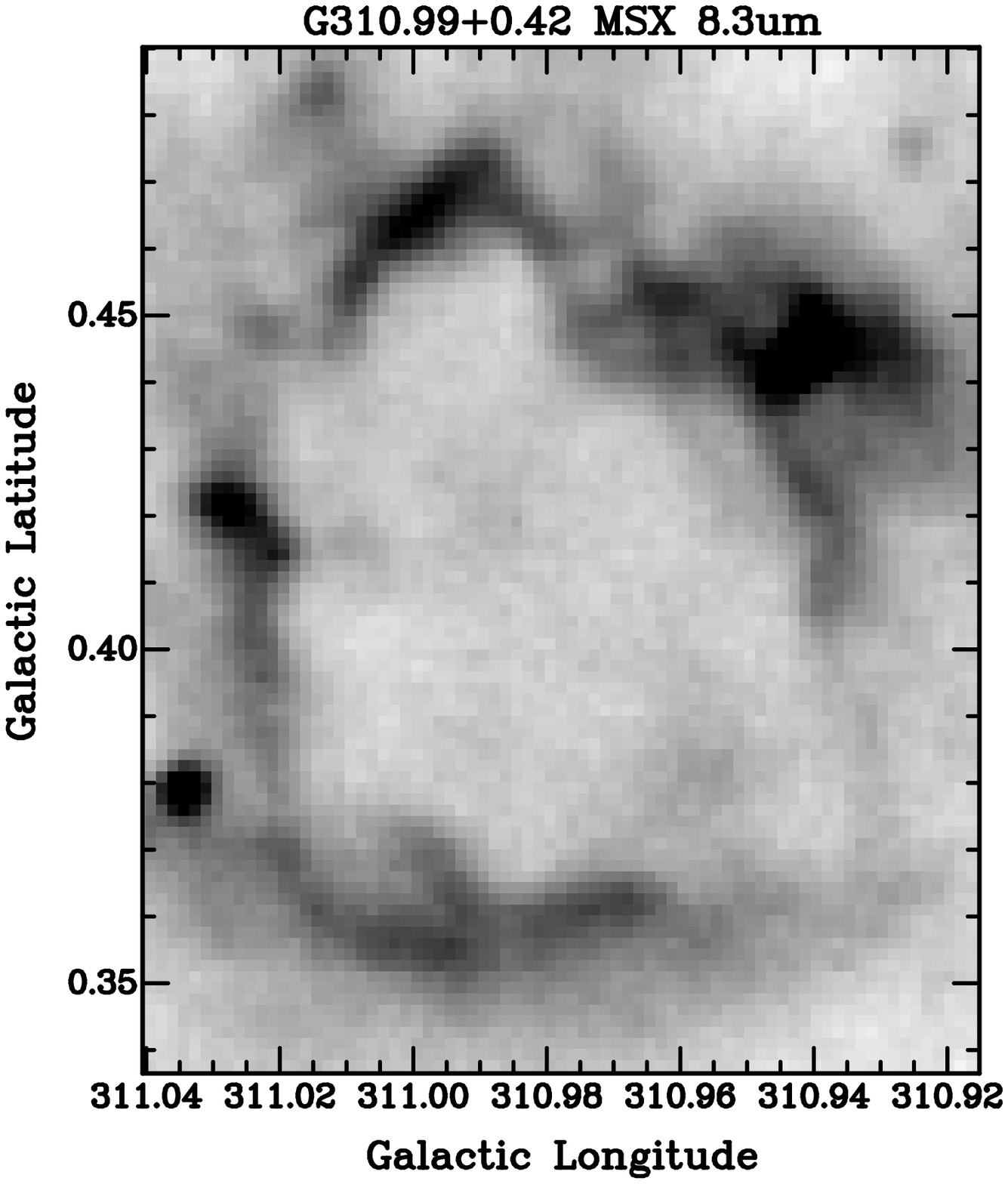}
\includegraphics{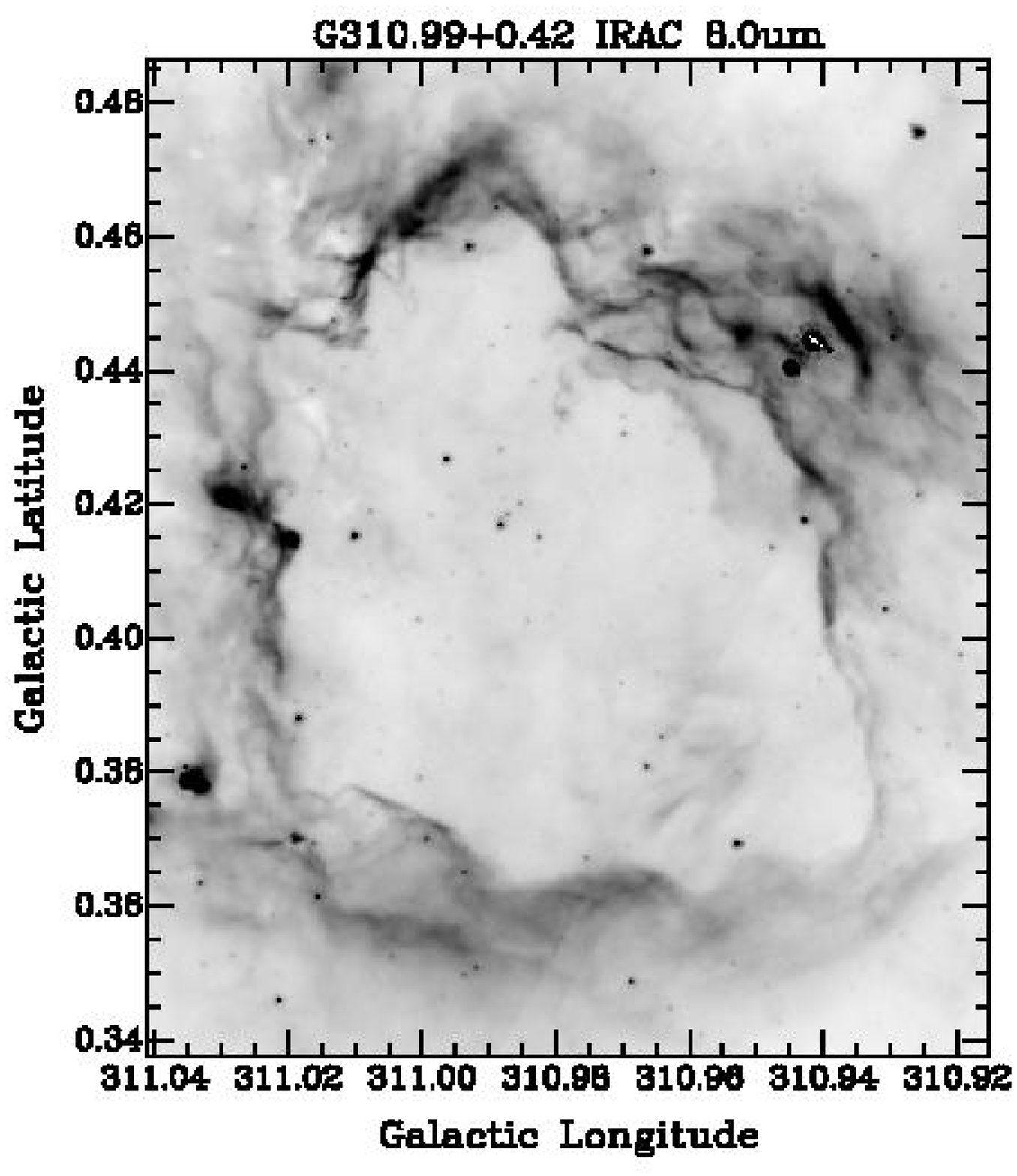}
\caption{Matched-area images of G310.99+0.42.  Left: MSX, 8.3\,$\mu$m.  
Right: {\it IRAC}, 8.0\,$\mu$m.}
\label{mira1}
\end{figure}

\begin{figure}
\vspace{9.5cm}
\includegraphics{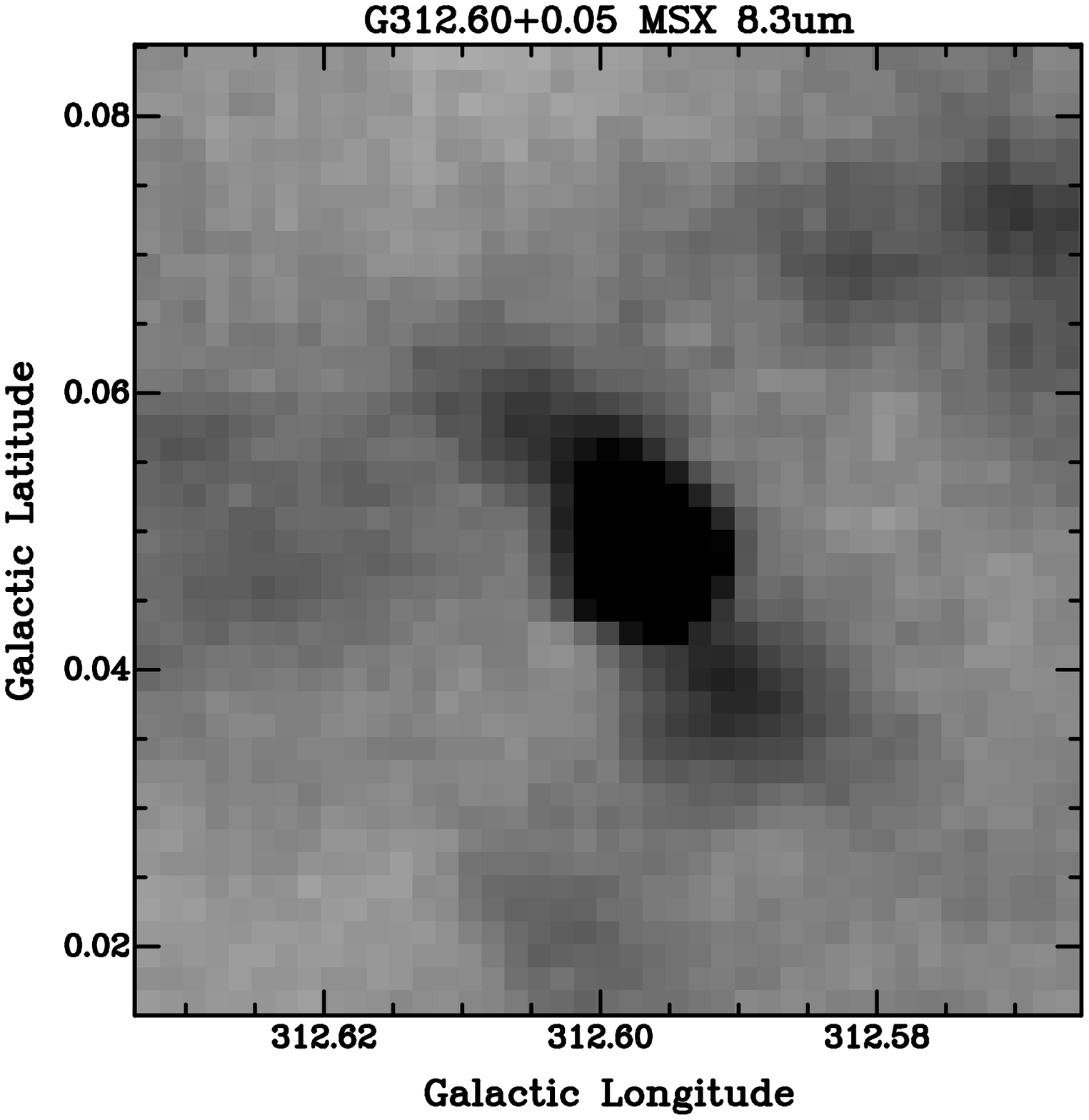}
\includegraphics{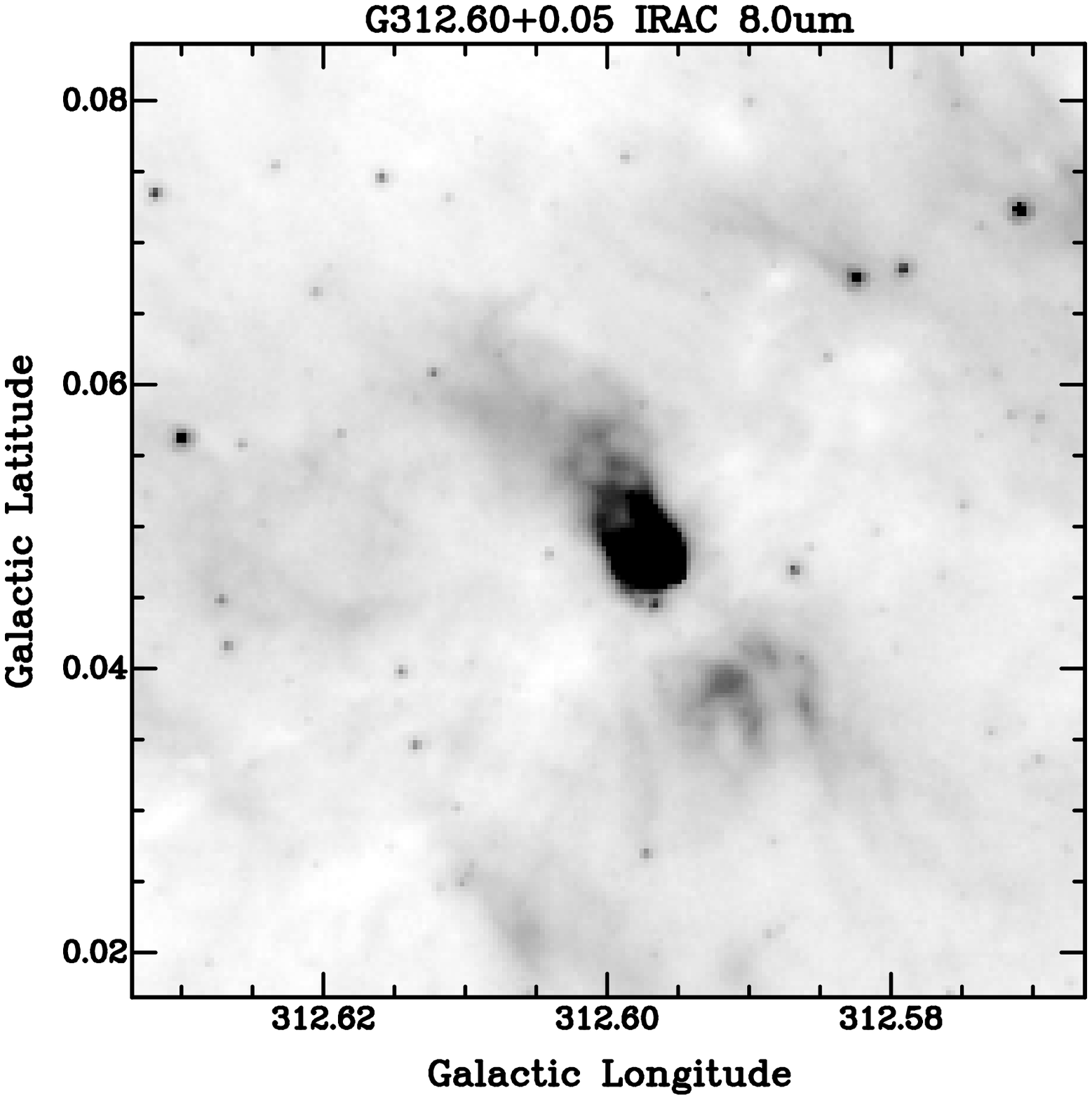}
\caption{Matched-area images of G312.60+0.50.  Left: MSX, 8.3\,$\mu$m.  
Right: {\it IRAC}, 8.0\,$\mu$m.}
\label{mira2}
\end{figure}

\twocolumn
\noindent
a 12\,percent uncertainty to the {\it IRAC} spatial flux
integrals.  The associated tests involved varying the choice of local areas from 
which to assess the diffuse sky background.

\subsection{Direct comparison of MSX and {\it IRAC} images}
To illustrate the similarities and differences between MSX and {\it IRAC} images of the
same regions we present grey-scale images of two fields (Figs.~\ref{mira1} and \ref{mira2}).  
MSX images of H{\sc ii} regions are generally sparsely populated by
point sources.  We have chosen G310.99+0.42 because it shows a rather typical
situation: many point sources but almost all of them undetected by MSX. The
few that MSX saw are mostly avoidable by careful selection of the bounding
rectangle for integration, or are readily isolated and integrated separately 
(which we generally did).  By contrast we also show a compact region apparently 
immersed in diffuse emission: G312.60+0.50, where SST shows a wealth of point 
sources but MSX does not.

\begin{figure}
\vspace{7.0cm}
\includegraphics{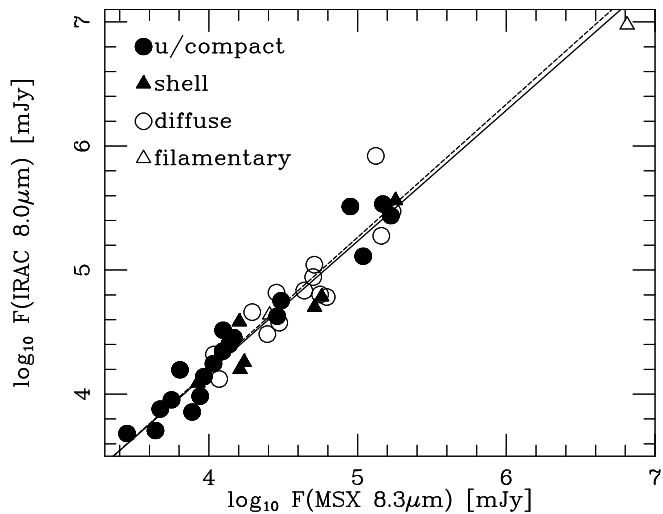}
\caption{Log-log plot of {\it IRAC}\,8.0-$\mu$m vs. MSX\,8.3-$\mu$m flux densities distinguishing
H{\sc ii} regions of four different characters: ultracompact and compact; shells; diffuse;
and filamentary.  The key indicates the symbol associated with the radio character of each 
object, based on our new MIR morphology. The two lines represent the regression
relationships for 42 regions (dashed) and for the subsample of the 25 apparently 
smallest and densest sources (solid). }
\label{sstmsx}
\end{figure}

The fifth column in Table~1 gives the ratios of {\it IRAC}\,8.0-$\mu$m to MSX\,8.3-$\mu$m
integrated fluxes, with their statistical uncertainties.  Fig.~\ref{sstmsx} presents the
regression of these MIR fluxes, with errors in both variables.  Symbols distinguish between 
H{\sc ii} regions of four different characters as indicated in the key (ultracompact and compact;
shells; diffuse; and filamentary.)  The two lines differentiate between the regressions for 
the total sample of 42 regions available with the SST (dashed) and for the
subsample of the 25 regions with the best defined morphologies (solid); i.e. for
all objects that we categorize as ultracompact, compact, or shells.  The unusual object, 
G311.70+0.31 (\S9), was included with these compact sources because of its small apparent size
and well-defined structure. This subsample lacks all the diffuse and filamentary regions.  

Table~2 summarizes these results for each ratio of attributes by subgroup, 
giving the median, standard error of the median, size of the subsample, slope and offset with 
$1\sigma$ errors for the regression between logarithmic quantities, and the Pearson
correlation coefficient between the two sets of MIR flux densities.  The table is essentially
divided between those regions that are well-defined and/or dense, those with complete or
partial shell-like structures, and those consisting primarily of diffuse material, 
tenuous filaments, or bubbles.  Specifically the categories of H{\sc ii} regions for which 
results are presented in Table~2 are: ultracompact and compact combined (``U/C''); shells (``S''); 
ultracompact, compact and shells combined (``U/C/S''); diffuse and filamentary combined
(we have only two filamentary regions) (``D/F''); and the entire sample. 
Further conbinations of morphological subcategories reveals no meaningful differences in the ratio
IRAC/MSX with respect to the character of the region.

The regression for the entire sample has a slope of 1.07$\pm$0.03
(1.00$-$1.14, $\pm3\sigma$ range), while that for the 25 densest and
apparently smallest objects has a slope of 1.05$\pm$0.03 (0.98$-$1.13, $\pm3\sigma$ range).
There is no statistically significant difference between the two correlations.  Both are consistent with 
linear proportionality between SST and MSX 8-$\mu$m fluxes, have good correlation coefficients, 
and span three and a half orders of magnitude in observed dynamic range of MIR flux.

We have also carried out an independent
analysis of the data offered in this paper using the types assigned by WCL (cometary, compact, 
classical, ``H{\sc ii}", ``thermal", filamentary (including bubbles), low-density).  These
are somewhat informal designations drawn from heterogeneous literature and there
are no significant distinctions in terms of the ratio of {\it IRAC} to MSX 8-$\mu$m flux densities
even among WCL's types.

The entire sample available to define the ratio of {\it IRAC}/MSX near~8$\mu$m consists of 42 regions 
and yields a median of about 1.6.  There are no significant differences between this
and any subsample of regions.  The similarity of the ratios in
physically very different sources such as diffuse and compact H{\sc ii} regions
confirms the reality of this effect and marks it as a problem with the {\it IRAC} calibration.  
Our best quantitative estimate of the effect comes from an independent median analysis of the 
25 smallest and densest regions: 1.55$\pm$0.15.  This is the empirical ratio of
IRAC\,8.0-$\mu$m/MSX\,8.3-$\mu$m found for these regions.

\begin{sidewaystable*}
\centering
{\bf Table~2}. Analysis of ratios of {\it IRAC}\,8.0-$\mu$m/MSX\,8.3$\mu$m, 
MSX/MOST, and {\it IRAC}/MOST fluxes and sample sizes (in parentheses), 
and log-log regressions.
\vspace*{4mm}
\small
\begin{tabular}{rrrrrcrrrrcrrrr}
\cline{1-15}\\
 & \multicolumn{4}{c}{IRAC/MSX}& &   \multicolumn{4}{c}{MSX/MOST}& & \multicolumn{4}{c}{IRAC/MOST}\\
\cline{2-5} \cline{7-10} \cline{12-15}
Group&  median$\pm$&      slope$\pm$&        offset$\pm$&      correl.&
   &   median$\pm$&      slope$\pm$&        offset$\pm$&      correl.& 
   &   median$\pm$&      slope$\pm$&        offset$\pm$&      correl.\\
   &  sem(no.)& $\sigma$& $\sigma$& coeff.&
   &  sem(no.)& $\sigma$& $\sigma$& coeff.&
   &  sem(no.)& $\sigma$& $\sigma$& coeff.\\
\cline{1-15}\\
U/C&   1.6$\pm$0.2(19)&  1.07$\pm$0.03& $-$0.1$\pm$0.1& 0.97& 
&          25$\pm$5(18)&     1.00$\pm$0.02& 1.3$\pm$0.1& 0.88&
&          36$\pm$13(18)&    0.98$\pm$0.03& 1.6$\pm$0.1& 0.82\\
S&         1.5$\pm$0.3(6)&  1.14$\pm$0.07& $-$0.5$\pm$0.3& 0.96& 
&         45$\pm$20(6)&    1.11$\pm$0.05& 1.5$\pm$0.1& 0.87&
&         78$\pm$24(6)&    1.21$\pm$0.07& 1.3$\pm$0.2& 0.84\\
U/C/S& 1.6$\pm$0.2(25)&  1.05$\pm$0.03& $-$0.1$\pm$0.1& 0.97& 
&         27$\pm$13(24)&     0.99$\pm$0.01& 1.4$\pm$0.1& 0.83&
&         41$\pm$14(24)&    0.93$\pm$0.02& 1.8$\pm$0.1& 0.79\\
D/F&   1.6$\pm$0.4(17)&  1.11$\pm$0.04& $-$0.3$\pm$0.2& 0.96& 
&         44$\pm$18(18)&    0.77$\pm$0.03& 2.3$\pm$0.1& 0.72&
&         82$\pm$25(17)&    1.29$\pm$0.04& 0.9$\pm$0.1& 0.80\\
All&   1.6$\pm$0.2(42)&  1.07$\pm$0.03& $-$0.1$\pm$0.1& 0.97& 
&         33$\pm$18(43)&    0.93$\pm$0.02& 1.6$\pm$0.1& 0.77&
&         63$\pm$26(42)&    1.03$\pm$0.02& 1.7$\pm$0.1& 0.77\\
\cline{1-15}\\
\end{tabular}
\end{sidewaystable*}

\section{MIR spectra of H{\sc ii} regions: the diffuse calibration of {\it IRAC} at 8.0~microns}
In addition to the SSC's formal study of aperture correction factors, there is an 
analysis of the {\it IRAC} extended source calibration by Jarrett (2006), now an official SSC 
page\footnote{http://ssc.spitzer.caltech.edu/irac/calib/extcal/},
based on a sample of twelve elliptical galaxies whose observed light distributions have been modeled.
That work finds a flux correction factor that varies from 0.78 to 0.74 at 8.0\,$\mu$m, with a 
probable uncertainty of 10\,percent.
This factor is the inverse of our {\it IRAC}/MSX ratio and, therefore, implies ratios
of 1.28$-$1.35, adopting MSX as the fiducial.  That the ratios of {\it IRAC}/MSX exceed unity
is believed to be due to scattering of light internal to the {\it IRAC} detector arrays, and
it is a particular problem in the Si:As arrays (the 5.8 and 8.0-$\mu$m bands).  The 
origin of the scattered light is thought to be in the epoxy layer between the detector and 
multiplexer.  The glue has a strong spectral response, peaking in the 5$-$6-$\mu$m 
region (Hora et al. 2004a).  

\begin{figure}
\vspace{6.5cm}
\includegraphics{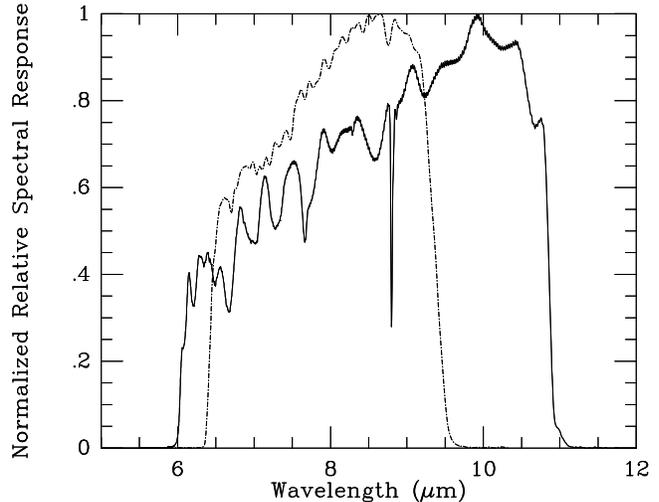}
\caption{Comparison of the in-band portions of the relative spectral response curves 
for MSX's 8.3-$\mu$m (solid) and {\it IRAC}'s 8.0-$\mu$m bands (dash-dot).}
\label{8umcf}
\end{figure}

Not all of this excess is attributable to light scattering, however.  The {\it IRAC}\,8.0-$\mu$m
band has a very different relative response curve from the broader, somewhat longer,
MSX\,8.3-$\mu$m band (Fig.~\ref{8umcf}), and the effects of this difference when integrating over the spectra 
of H{\sc ii} regions must be treated. This we have done by three different methods.  First 
we used the 2$-$35-$\mu$m template for H{\sc ii}
regions, one of 87 such templates developed by Cohen (1993) to give the SKY model of 
the point source sky its capability to predict source counts for arbitrary IR bands.
This normalized template spectrum was constructed as the average of the observed 
spectra of 10
bright, rather compact (measured with beams between 3.4$^{\prime\prime}$ and 30$^{\prime\prime}$ in diameter) 
H{\sc ii} regions.  Within these {\it IRAC} and MSX bands the dominant spectral features from the cores 
($\sim$1~arcmin in diameter) of H{\sc ii} regions are (ranked by strength): a strong dust 
continuum, very deep 10-$\mu$m silicate absorption; deep 6.0-$\mu$m water absorption; PAH 
emission at 7.7\,$\mu$m; and, sometimes, weak fine structure lines (Willner et al. 1982).
Integrating the two spaceborne bands over this ($\lambda$,F$_\lambda$) spectrum yields 
in-band fluxes of 1.75$\times10^{-15}$~W~cm$^{-2}$ (SST) and 1.90$\times10^{-15}$~W~cm$^{-2}$
(MSX).  Dividing by the corresponding bandwidths ($1.146\times10^{13}$ and
$1.402\times10^{13}$~Hz) gives 0.0153 and 0.0135~Jy, respectively.  Consequently, in
an optically perfect instrument, one would expect 
{\it IRAC}/MSX~=~1.12$\pm$0.03. Away from the embedded central ionizing 
source, the spectrum of extended emission from the PDR is still that of cool 
dust, overlaid by spatially extended silicate absorption, with diffuse PAH emission, 
and fine structure lines, even in large regions such as M17 (e.g. Kassis et al. 2002).
Therefore, Cohen's template is still relevant because the bulk of the energy is
emitted from the hot central core of an extended H{\sc ii} region. 

The second method seeks to define the best average spectrum of the H{\sc ii} regions
in the field under study here, rather than relying on an average composed of the most 
famous and brightest compact regions.  The H{\sc ii} region template is a combination
of ground-based 

\newpage
\onecolumn
\begin{figure}
\vspace{22cm}
\includegraphics{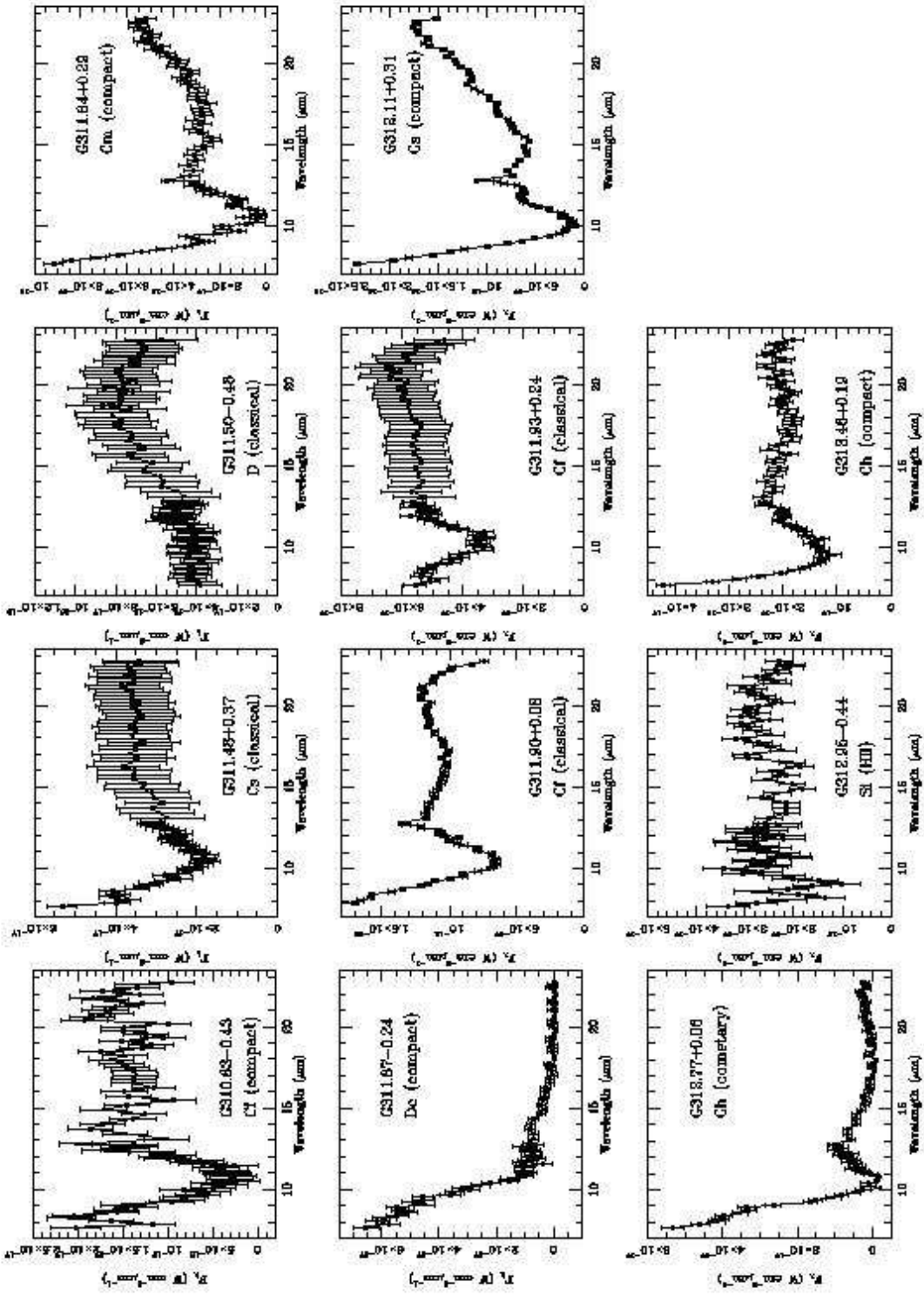} 
\caption{Calibrated LRS spectra for 11 H{\sc ii} regions in the field.  Error bars are
$\pm1\sigma$ at each wavelength.  Our own, and WCL's (in parentheses), morphological 
types are given in each separate plot.}
\label{lrs}
\end{figure}
\newpage

\twocolumn
\noindent
and airborne spectra, bridging gaps using data from the {\it IRAS} Low Resolution 
Spectrometer (LRS: Wildeman, Beintema \& Wesselius 1983).  LRS coverage is from
7.7$-$22.7\,$\mu$m and each spectrum consists of two sections, blue and red, whose
effective apertures were $6^{\prime}\times5^{\prime}$ and $6^{\prime}\times7.5^{\prime}$,
respectively.  For substantially extended sources the convolution 
of spatial structure and wavelength in these large apertures requires a complex extraction
process.  Sources smaller than 15$^{\prime\prime}$ are not affected by the convolution
(Assendorp et al. 1995).  However, many H{\sc ii} region spectra were published in the
Atlas of LRS spectra (Olnon et al. 1986), based on a much simpler extraction
method.  This approach yielded perfectly adequate spectra (benchmarked by Cohen (1993)
against ground-based and airborne spectra in the construction of SKY's 87 templates)
because most of the bright infrared emission of H{\sc ii} regions arises in
volumes of small apparent size.  The complete Dutch LRS database comprised 171,000
LRS spectra, derived by using the same
software as Olnon et al. (1986), and we have sought spectra for 
all the H{\sc ii} regions in the {\it l}~=~312$^\circ$ field in this database.
We incorporate the recalibration of LRS spectral shape provided by Cohen et al. (1992a),
and normalize the resulting spectra to the absolute 12-$\mu$m flux densities (scaled
by the factor of 0.961 described by Cohen et al. (1996: their Table 6) as applicable to
IRAS 12-$\mu$m flux densities of both the Point Source Catalog (PSC) and the Faint Source 
Survey).  When available we used the PSC flux density; otherwise that from the Faint
Source Reject catalog. We found 93 LRS spectra for 32 regions, and were able to extract
useful averages for 11 of these regions.

Fig.~\ref{lrs} presents the 11 averaged, calibrated spectra.  Nine of these are somewhat
similarly shaped; two others are quite distinct 
because of their blueness: G311.87$-$0.24 (Dc) and the cometary G312.77+0.06 
(Ch).  Each individual spectrum is plotted in F$_\lambda$, and carries 
the source designation, and both our type and that assigned by WCL.  Error bars are $\pm1\sigma$ at 
each wavelength.  Quality varies from source to source although the characteristic deep
10-$\mu$m silicate absorption is seen in almost all the objects.  Deep silicate absorptions
might indicate the presence of accretion disks that are viewed edge-on, but the frequency of 
these in H{\sc ii} regions suggests that, in most regions, these absorptions occur because of 
deep embedding of the exciting stars in more spheroidal dusty envelopes.

\begin{figure}
\vspace{7.0cm}
\includegraphics{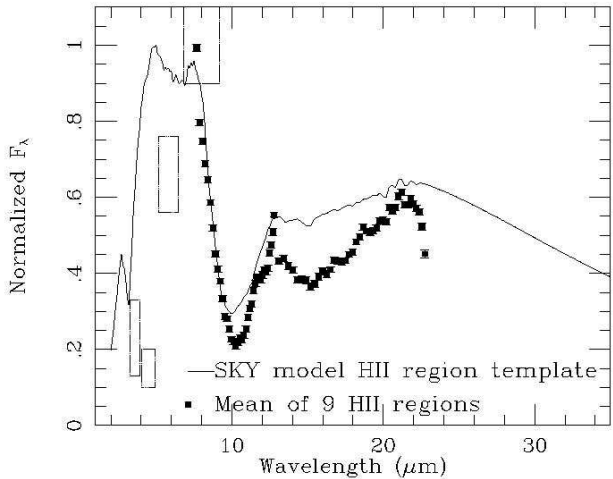}
\caption{Comparison between the SKY template for generic H{\sc ii} regions (solid line)
and the average of 9 regions in this field (filled squares with$\pm1\sigma$ error bars).
Open rectangles show the locations of the normalized median SED for the regions (\S6).}
\label{skycfn}
\end{figure}

We created the average of the nine most similar spectra by inverse-variance weighting each 
spectrum (peak normalized to unity) based on the wavelength-dependent uncertainties that 
our reduction code produces (Cohen et al. 1992a).  Fig.~\ref{skycfn} compares this 
average LRS spectrum with the complete 2$-$35-$\mu$m spectral template from the SKY model.  
The agreement between the two normalized spectra is very good between the LRS cut-on (7.7\,$\mu$m)
and the long wave limits of the bandpasses of the relevant MSX (e.g., 1\,percent 
transmission at 11.08\,$\mu$m) and SST filters (1\,percent transmission at 
9.63\,$\mu$m). In the MIR, ionic lines do not contribute greatly to the 
content of the SST and MSX bands.

We have constructed a hybrid spectrum by taking the data below 7.67\,$\mu$m 
from the SKY template, and the average of the nine LRS spectra for our 
actual H{\sc ii} regions beyond this wavelength.  We integrated the 
MSX\,8.3-$\mu$m and SST 8.0-$\mu$m relative spectral response curves over
this spectrum, resulting in a ratio of the two instruments' flux densities of {\it IRAC}/MSX~=~1.17$\pm$0.03.  

For the truly diffuse ISM
the same calculations can be performed using spectra taken in the large-aperture 
($8\times8$~arcmin) near-infrared and MIR 
spectrometers of the Infrared Telescope in Space (IRTS: Onaka 2003, Sakon et al. 2004).  
We have assembled a spectrum from 1.4$-$11.67\,$\mu$m based on the IRTS and for this the 
ratio of integrated flux densities, {\it IRAC}/MSX, is 1.15$\pm$0.06.  The IRTS spectrometers 
were calibrated (Tanabe et al. 1997, Onaka et al. 2003, Murakami et al. 2004)
with the identical basis to that embracing the {\it IRAS} LRS, MSX, and the 
SST so that integrations over this diffuse spectrum are directly comparable to those 
over the LRS and SKY template spectra.  The same procedure can be 
applied to the ISOCAM CVF spectra of the ISM presented by Flagey et al. (2006).  We 
computed the ratio {\it IRAC}/MSX for their spectrum with the most relevant longitude, the
299.7$^\circ$ field, and obtained 1.11.  However, the traceability of the 
CAM CVF flux calibration is not identical to that of the MSX and SST and we assign this 
value twice the error of that derived from the IRTS spectrum.
Consequently, whether we observe the most compact of cores, spatially extended MIR emission 
associated with H{\sc ii} regions, or the ISM at large, we obtain a robust result for the
measured ratio of {\it IRAC}/MSX near 8\,$\mu$m with the empirical spectra of diffuse H{\sc ii} regions,
namely 1.14$\pm$0.02 (by inverse-variance weighting of our four values).  

The actual ratio
determined for {\it IRAC}/MSX is 1.55, of which the factor of 1.14 is attributable solely to the
different in-band fluxes measured by these two bands for the MIR spectrum of H{\sc ii} regions. 
Internal scattering in {\it IRAC}'s 8.0-$\mu$m detector, therefore, contributes the 
additional factor of 1.36$\pm$0.13 to the ratio of {\it IRAC}\,8.0-$\mu$m/MSX\,8.3-$\mu$m based on our
analysis of H{\sc ii} regions.  Expressed as the reciprocal quantity (the aperture or 
flux correction factor of SSC and Jarrett (2006)) our analysis is equivalent to 0.74$\pm$0.07.  
For comparison, the average correction factor and a conservative uncertainty of the elliptical
galaxies at 8.0\,$\mu$m is 0.74$\pm$0.07, while Reach et al. (2005) recommend 0.737 
(no uncertainty specified) at 8.0\,$\mu$m.
Therefore, our independent analysis of the diffuse flux correction of {\it IRAC} at 8.0\,$\mu$m 
based on H{\sc ii} regions is consistent with two separate analyses based either on the 
performance of the instrument, or on models of the IR light distributions of elliptical galaxies.

In the low surface brightness regime, the minimum detectable diffuse emission in this field 
has an intensity at 8.3\,$\mu$m of about 1.0$\times10^{-6}$ W m$^{-2}$~sr$^{-1}$.  At the same 
location the {\it IRAC}\,8.0-$\mu$m level is 1.4$\times10^{-6}$ W m$^{-2}$ sr$^{-1}$, 
consistent with the same factor for {\it IRAC}/MSX as measured for the bright H{\sc ii} regions.

\section{Spectral energy distributions}
We have taken the spatially integrated flux densities (in Jy) from the images
and multiplied by the bandwidths (in Hz) to generate in-band fluxes.  These we converted to 
Vega-based magnitudes and, subsequently, into flux densities expressed in F$_\lambda$ 
(in W~cm$^{-2}$\,$\mu$m$^{-1}$, after dividing by the bandwidths in\,$\mu$m.  
Table~3 presents the absolute calibration for {\it IRAC} based upon the newest released relative spectral
response curves (July 2004) and the absolute Vega spectrum used to define zero magnitude (Cohen et 
al. 1992b) for {\it IRAC} and absolutely validated to 1.1\,percent by Price et al. (2004).  The
assumed uncertainty on the response curves is 5 percent, independent of wavelength.

\begin{table*}
\begin{center}
{\bf Table 3}. {\it IRAC} zero magnitude absolute attributes with isophotal quantities and uncertainties below 
\vspace{3mm}
\begin{tabular}{lcccccccc}
Filter&  Bandwidth&  In-Band& F$_\lambda$(iso)& $\lambda$(iso)&   Bandwdth&  F$_\nu$(iso)&  $\nu$(iso)\\
      &    \,$\mu$m&   W~cm$^{-2}$&   W~cm$^{-2}$\,$\mu$m$^{-1}$&    \,$\mu$m&     Hz&    Jy&         Hz\\
      &     Error& Error& Error&  Error&   Error&   Error&  Error\\
      &     \,$\mu$m&    \%&     W~cm$^{-2}$\,$\mu$m$^{-1}$&    \,$\mu$m&        Hz&        Jy&        Hz\\
\hline
IRAC1& 6.514E-01& 4.277E-15& 6.566E-15& 3.550E+00& 1.543E+13& 2.771E+02& 8.461E+13\\
 &     3.890E-03& 1.572E+00& 1.104E-16& 1.533E-02& 6.375E+10& 4.356E+00& 7.047E+11\\
IRAC2& 8.827E-01& 2.342E-15& 2.654E-15& 4.494E+00& 1.306E+13& 1.794E+02& 6.685E+13\\ 
 &     5.707E-03& 1.592E+00& 4.560E-17& 1.994E-02& 5.834E+10& 2.853E+00& 5.657E+11\\
IRAC3& 1.196E+00& 1.237E-15& 1.034E-15& 5.725E+00& 1.086E+13& 1.139E+02& 5.255E+13\\
 &     8.376E-03& 1.607E+00& 1.813E-17& 2.560E-02& 5.182E+10& 1.829E+00& 4.381E+11\\
IRAC4& 2.395E+00& 7.230E-16& 3.019E-16& 7.837E+00& 1.146E+13& 6.310E+01& 3.865E+13\\ 
 &     1.663E-02& 1.599E+00& 5.263E-18& 3.528E-02& 5.242E+10& 1.006E+00& 3.172E+11\\
\end{tabular}
\end{center}
\end{table*}

\begin{figure}
\vspace{7.0cm}
\includegraphics{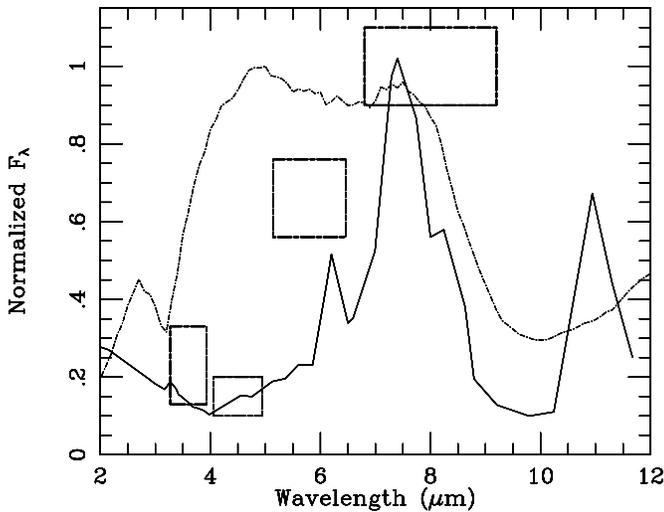}
\caption{Comparison of the median SED ($\pm$2 standard error rectangles) for thermal sources
in this field with the diffuse ISM (solid line: IRTS, \S4) and the SKY H{\sc ii} region template 
spectrum (short dashed line).}
\label{diffcfn}
\end{figure}

It is necessary, however, to apply corrections for what is known about the shortcomings
of the diffuse calibration of this instrument.  We have determined in this paper that the
8.0-$\mu$m calibration leads to fluxes too high by 36\,percent.  This can be compensated by
applying an equivalent offset to the diffuse magnitude scale of +0.33.  For {\it IRAC}'s 
three other bands we adopt the SSC analysis (footnote in \S5) based on model template 
SEDs for elliptical galaxies.  Averaging the ``correction factors" derived from 
these galaxies indicates that corresponding offsets of +0.19, +0.09, +0.33
should be applied to measured diffuse magnitudes at 3.6\,$\mu$m, 4.5\,$\mu$m, and 
5.8\,$\mu$m, respectively.  Table~4 incorporates these corrections for each band both to magnitudes
and to the values of F$_\lambda$.  The table summarizes integrated magnitudes of the sample
of H{\sc ii} regions giving the brightest and faintest magnitudes, and the medians with
their standard errors.  The dynamic range of the sample varies from a
factor of 380 at 3.6\,$\mu$m to 200 at 8.0\,$\mu$m, and the difference between the median
magnitudes at these two wavelengths (almost 5~mag) confirms the great redness of 
H{\sc ii} regions.  Every H{\sc ii} region has its maximum {\it IRAC} F$_\lambda$ in the 8.0\,$\mu$m band.
We normalized the quartet of F$_\lambda$ values for every object by dividing by
F$_{\lambda}$(8.0\,$\mu$m), then took the median of each complete set of normalized
values of F$_{\lambda}$(band)/F$_{\lambda}$(8.0\,$\mu$m) where band~=~3.6, 4.5, 5.8\,$\mu$m.
From the corrected magnitudes of the individual regions we have constructed Table~5, 
which presents the median values of the six colour indices between the four {\it IRAC} bands.
Note that uncertainties are given for each offset based on the standard deviation of
the average value of the colour's correction factors for the nine galaxies.  The errors in
corrected colours are the RSS of the errors in the corresponding offsets and the standard
errors of the medians.
Those medians, with their standard errors, appear in col.(6) of Table~4.  These
normalized values represent a coarse SED over the {\it IRAC} range for H{\sc ii} regions.
Fig.~\ref{diffcfn} directly compares this SED with the template for the cores of bright 
H{\sc ii} regions and the 
diffuse ISM template.  The SED is represented by rectangles, horizontally as
wide as each bandwidth and vertically as tall as $\pm$2 standard errors of the median.
There is obviously much better agreement with diffuse emission than with the spectrum
of a compact H{\sc ii} region with its exciting star(s) and hot dust continuum.

\begin{table*}
\begin{center}
{\bf Table 4}. Vega-based magnitudes of the sample of H{\sc ii} regions in the {\it IRAC} bands,
and normalized F$_\lambda$ SED.
\vspace{3mm}
\begin{tabular}{llrrcc}
Band& $\lambda$($\mu$m)& Brightest& Faintest& Median$\pm$s.e.med.& Median F$_{\lambda}$(band)/F(8.0\,$\mu$m)\\
\hline   
1&   3.6& 2.24& 8.68&     5.60$\pm$0.25&  0.23$\pm$0.05\\ 
2&   4.5& 2.02& 8.17&     5.32$\pm$0.25&  0.15$\pm$0.02\\ 
3&   5.8& $-$0.58& 5.21&  2.71$\pm$0.25&  0.66$\pm$0.02\\ 
4&   8.0& $-$2.47& 3.42&  0.54$\pm$0.25&  1.00$\pm$0.04\\ 
\end{tabular}
\end{center}
\end{table*}

\begin{table*}
\begin{center}
{\bf Table~5}. Observed and corrected median Vega-based colours (in magnitudes) of H{\sc ii} regions in 
the four {\it IRAC} bands
\vspace{3mm}
\begin{tabular}{cccc}
Colour&   Observed& Offset$\pm\sigma$& Corrected\\ 
&        median&  & median$\pm$s.e.med.\\ 
\hline
$[3.6]-[4.5]$&    0.43&       +0.11$\pm$0.02&  0.53$\pm$0.06\\ 
$[3.6]-[5.8]$&    3.35&     $-$0.13$\pm$0.03&  3.22$\pm$0.10\\    
$[3.6]-[8.0]$&    5.07&     $-$0.13$\pm$0.03&  4.94$\pm$0.10\\
$[4.5]-[5.8]$&    2.91&     $-$0.24$\pm$0.03&  2.67$\pm$0.07\\
$[4.5]-[8.0]$&    4.69&     $-$0.28$\pm$0.03&  4.41$\pm$0.08\\
$[5.8]-[8.0]$&    1.79&       +0.00$\pm$0.04&  1.79$\pm$0.05\\
\end{tabular}
\end{center}
\end{table*}

\begin{figure}
\vspace{8.5cm} 
\includegraphics{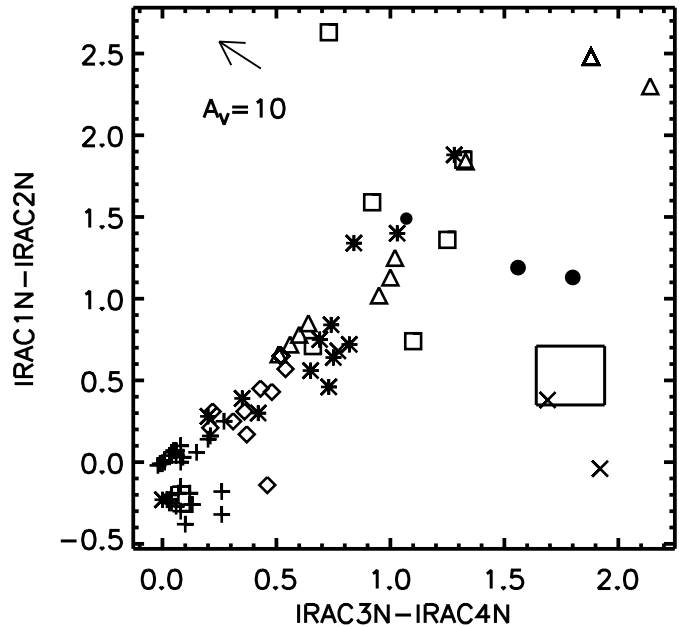}
\caption{Diagnostic {\it IRAC} colour-colour plot ([3.6]-[4.5] vs. [5.8]-[8.0])
comprising 87 types of IR point source. Key:
pluses - normal dwarfs, giants, supergiants; asterisks - AGB M stars; diamonds - AGB 
visible C stars; triangles - AGB deeply embedded IR C stars; squares - hyperluminous 
objects (these objects include deeply embedded OH/IR stars and early-type hypergiants (Cohen 1993;
a small number are required to reproduce MIR source counts at low latitude: Wainscoat et al. 
(1992)); crosses - exotica (T Tau stars, reflection nebulae); larger filled circles - 
planetary nebulae; small filled circle - bright compact H{\sc ii} regions.
The reddening vector corresponding to an A$_V$ of 10 mag is shown by the shaft 
of the arrow in the upper left corner.  The large rectangle denotes the zone 
occupied by the extended H{\sc ii} regions in this paper ($\pm$3 standard errors). 
Axis labels include magnitudes with ``N" in their names to signify that new 
relative spectral response files from July 2004 were used for the {\it IRAC} bands.}
\label{1234col}
\end{figure}

\begin{figure}
\vspace{8.5cm}
\includegraphics{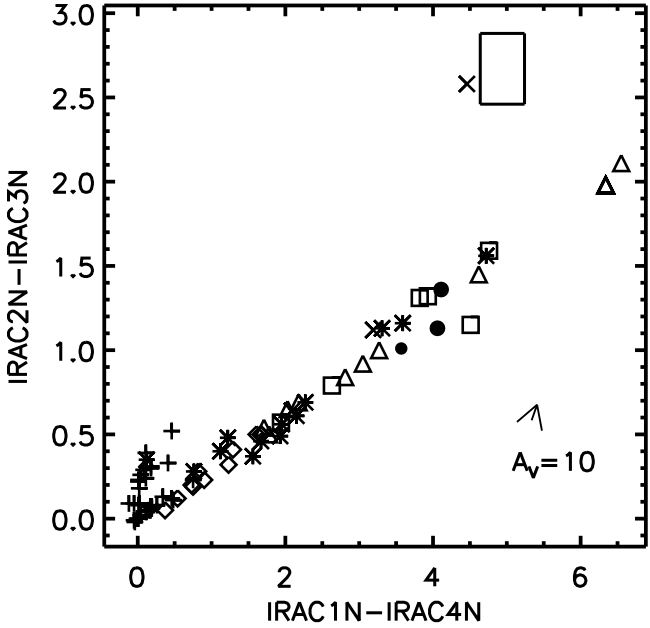}
\caption{As Fig.~\ref{1234col} but for [4.5]-[5.8] vs. [3.6]-[8.0].}
\label{2314col}
\end{figure}

Figures~\ref{1234col} and \ref{2314col} present two interpretive colour-colour planes 
for {\it IRAC} bands showing the 87 categories of source in the version of the SKY model 
where each source is represented by a complete 2$-$35-$\mu$m spectrum (Cohen 1993).  
The caption to Fig.~\ref{1234col} details the symbols used for various populations.  
These colour-colour plots can be used for quick-look diagnostics of populations in
random Galactic directions.  Planetaries are represented by two large filled circles because
young and evolved planetary nebulae occupy two distinct regions in {\it IRAS} colour-colour 
planes (Walker \& Cohen 1988; Walker et al. 1989).  Hora et al. (2004b) provide empirical 
IRAC colour-colour locations from their study of planetary nebulae with SST and their
colours encompass the two larger circles predicted in Fig.~\ref{1234col}.  
 
\begin{figure}
\vspace{7.5cm}
\includegraphics{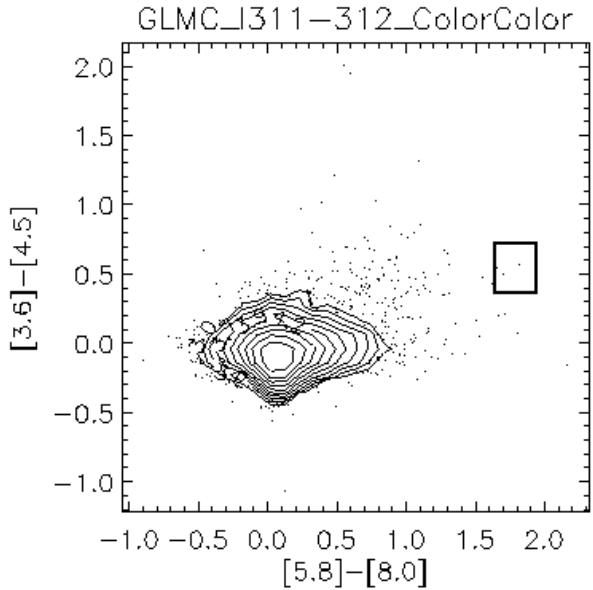} 
\caption{Same axes as in Fig.~\ref{2314col} but based on the 213563 observed GLIMPSE point 
sources in the catalogue for this field.  The densest regions of the plot are represented by 
contours rather than showing every source.  The appropriate H{\sc ii} region
box shown in Fig.~\ref{1234col} also appears in this plot.}
\label{glimpsecc}
\end{figure}

These figures show [3.6]-[4.5] vs. [5.8]-[8.0] and [4.5]-[5.8] vs. [3.6]-[8.0]
diagrams.  The large rectangles ($\pm$3 standard errors of the median) 
represent the zones in which diffuse H{\sc ii} regions ought to fall 
based on the median corrected colours in col.(4) of Table~5.  Correcting the observed
colours using the offsets in col.(3) of Table~5 results in changes that are smaller
than those that affect the magnitudes in any single band, and in no way alter the
almost complete isolation of the diffuse H{\sc ii} regions' box in these plots.  Only
one of the 87 types of point source in these plots encroaches on the diffuse colour box
(the cross, corresponding to one type of reflection nebula.)
Fig.~\ref{glimpsecc} displays the observed colour-colour plots for the GLIMPSE Catalog 
of point sources for the roughly 1~deg$^2$ field centred at {\it l}~=~312.50$^\circ$.  
Ultracompact sources would pass into the GLIMPSE Point Source Catalog for this field
provided that their cores were not saturated in any of the four bands nor contained any other
corrupted pixels within the core radius.  However, their colours would mimic the bright,
compact H{\sc ii} region colours represented by the small filled circle in each of 
Figs.~\ref{1234col} and \ref{2314col}.  It is interesting to note that there are three
such pointlike sources observed with the colours of (ultra)compact H{\sc ii} regions.
There is only trivial contamination, by any of the 213563 observed point sources in 
the GLIMPSE Catalog  for the 1\,deg$^2$ region centred on 312.50$^\circ$, of the 
corresponding colour-colour box that we have determined for diffuse H{\sc ii} regions.

The overall dispersion in SEDs expressed by the {\it IRAC} colours is quite small, in part
because of the rather limited contributions of emission lines in most H{\sc ii} regions.
The primary features in their MIR spectra are attributable to broad absorptions
by silicates, ices, and organic materials.
Cohen et al. (2005) considered the emission in the {\it IRAC} bands from ionized gas in a new
planetary nebula found in GLIMPSE observations.  Both free-free and hydrogen line emission
were derived from the level of thermal radio emission but did not amount to more than
a few percent in the {\it IRAC} bandpasses.  Flagey et al. (2006) perform the same kind
of estimates for the diffuse ISM, deriving no more than about a 12\,percent contribution
from recombination lines and free-free continuum even at the shorter {\it IRAC} wavelengths.

Fine structure lines are not bright in H{\sc ii} regions as evidenced by their 
absence even in well-known compact sources (e.g. Willner et al. 1982).
One can similarly gauge the minimal influence of emission lines in the LRS range
through Fig.~\ref{lrs}.  Even for the IR-brighter sources in our field one sees
only three 12.8-$\mu$m [NeII] lines and one object might show an 18.7-$\mu$m 
[SIII] line.

\section{COMPARISON OF MIR AND RADIO FLUX DENSITIES}  
\begin{figure}
\vspace{7.0cm}
\includegraphics{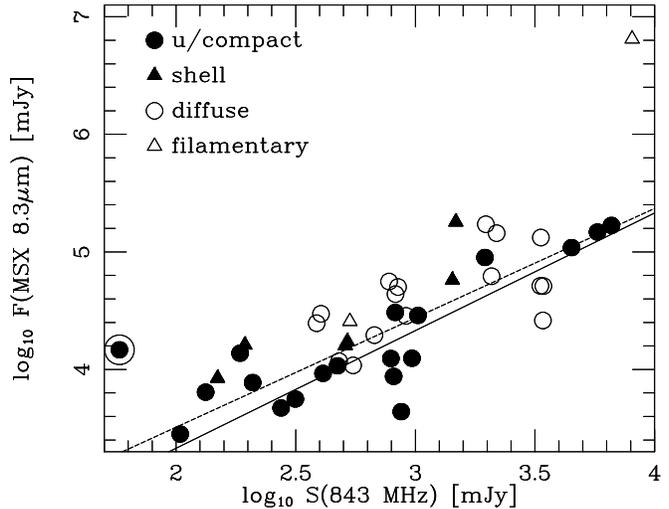}
\caption{Log-log plot of MSX\,8.3-$\mu$m vs. S(843~MHz) flux densities.  Symbols and
regression lines as in Fig.~\ref{sstmsx}. The compact object near the left boundary in
the plot whose filled symbol is also encircled represents the sole cometary region, 
G312.77+0.06, which has been excluded from the regression analyses of compact regions.} 
\label{hiimsx}
\end{figure}

\begin{figure}
\vspace{7.0cm}
\includegraphics{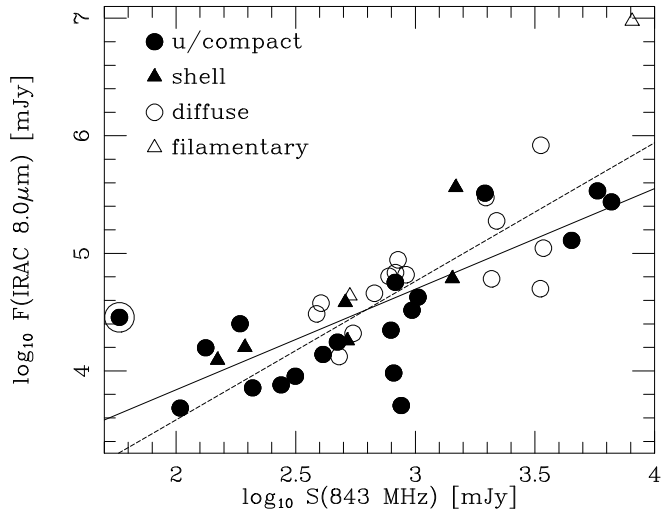}
\caption{Log-log plot of {\it IRAC}\,8.0-$\mu$m vs. S(843~MHz) flux densities.  Symbols and
regression lines as in Fig.~\ref{sstmsx}.}
\label{hiisst}
\end{figure}

Fig.~\ref{hiimsx} compares the MSX\,8.3-$\mu$m flux densities with those measured
from MOST images (both in mJy). Fig.~\ref{hiisst} similarly compares {\it IRAC}\,8.0-$\mu$m flux 
densities with 843-MHz data.  These two figures retain the different symbols to identify 
different types of thermal region. Table~2 summarizes the median analyses and regression 
lines for the subsamples of sources in terms of the ratios of MSX to MOST (centre of the 
table), and SST to MOST (right section of the table), flux densities.  

There are meaningful correlations implying linear proportionality between both MSX and 
SST 8-$\mu$m fluxes and radio continuum fluxes.  However, unlike the {\it IRAC}/MSX correlation, 
there are three issues which affect the MIR to radio flux ratios.
One expects the MOST to lose flux substantially as the interferometer
observes larger regions.  The best model for the actual synthesized beam of the MOST
incorporates an effective minimum spacing of about 70$\lambda$ (Hogan 1992) although this 
parameter varies between observations and may be dependent on meridian distance. Therefore,
angular scales greater than about 30\,arcmin are typically poorly detected.  This 
certainly affects the largest structure in the field (the filamentary object, G312.45+0.08, 
MSX/MOST of 800) with a size greater than 1$^\circ$. It might also overestimate MSX/MOST
for the second largest source (the diffuse G311.37+0.79, MSX/MOST of 87), with a size of 24\,arcmin,
corresponding to the second largest H{\sc ii} region in our whole sample. To assess the 
fraction of true radio continuum flux that has been measured for each specific region by the 
MOST would require detailed modeling based on the irregular structures of real
sources, and on the {\rm u-v} coverage obtained, like the simulations performed by Bock \& 
Gaensler (2005).  We can state only that MSX/MOST and {\it IRAC}/MOST ratios are invalid for 
G312.45+0.08 but we suggest that this phenomenon may also contribute to some of the other 
large ratios that are found in Table~2, particularly for diffuse regions.  

The second concern 
is relevant to the single example of a cometary H{\sc ii} region in our sample, G312.77+0.06.  
This appears anomalous in both Figs.~\ref{hiimsx} and \ref{hiisst} because it is quite 
compact yet also has a very large MIR/radio ratio.  We believe this object is optically 
thick at 843~MHz, greatly diminishing the radio continuum flux density.  
This would be consistent with the extremely deep 10-$\mu$m silicate absorption whose optical
depth, $\tau$(10\,$\mu$m)~$\approx$3, hence A$_V$$\approx$55 (Roche \& Aitken 1984;
Chiar \& Tielens 2006), and the fact that this source has the most extreme LRS energy 
distribution in our sample in that it shows the deepest 18-$\mu$m silicate absorption too.  
The blueness of the spectrum implies an abundance of warm ($\geq$400~K) dust (rather than 
cool, i.e. $\leq$100~K) local to the embedded high-mass star.  Consequently, this region's
exciting star is enormously embedded in dust, plausibly leading to local free-free 
absorption from the associated dense ionized gas.  Therefore, we have chosen to exclude
this region from any MIR/radio analysis of subsamples, reducing the ultracompact and compact
H{\sc ii} regions to 18 objects.  However, samples designated as ``All'' 
in Table~2 include this cometary region.

Thirdly, our SST images have been culled of point sources (only their residuals after PSF subtraction
remain) but we have no such products for MSX.  Thus MSX fluxes within the radio source rectangles
could potentially still include some faint point sources although efforts were made to remove their 
influence.  Radio flux densities for big sources will suffer loss of signal because
of resolving out structure as discussed above.  {\it Spitzer} images are
so sensitive to diffuse emission at low surface brightness levels that it could be necessary to
seek uncontaminated sky at greater distances from the source than for MSX.  This 
could lead to underestimated sky backgrounds, hence somewhat enlarged source fluxes for {\it IRAC},
and relatively reduced fluxes for MSX.  Both effects could inflate the ratio of {\it IRAC}/MOST.
Consequently, we chose matching sky regions for both {\it IRAC} and MSX to mitigate this problem.
Our best estimate for these ratios will clearly come from the sample of the 18 most compact 
entities, excluding the cometary source.  From this set we find MSX/MOST has a median of 25$\pm$5,
in excellent accord with CG's median of 24.  The corresponding empirical ratio (i.e. with no
flux correction factor applied) for SST is 36$\pm$13.  

\begin{figure}
\vspace{7.0cm}
\includegraphics{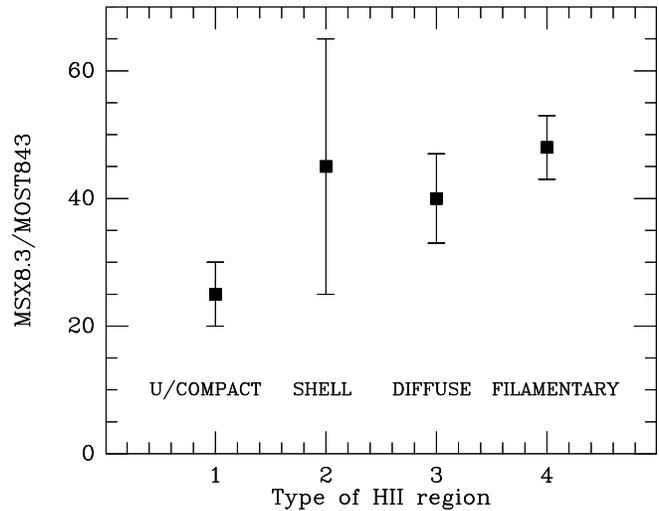}
\caption{Medians and their standard errors for the ratio of MSX\,8.3$\mu$m and MOST 
flux densities plotted sequentially for the four types of H{\sc ii}
region in the field.  Our MIR morphologies are indicated under each median.}
\label{msxmedtyp}
\end{figure}

\begin{figure}
\vspace{7.0cm}
\includegraphics{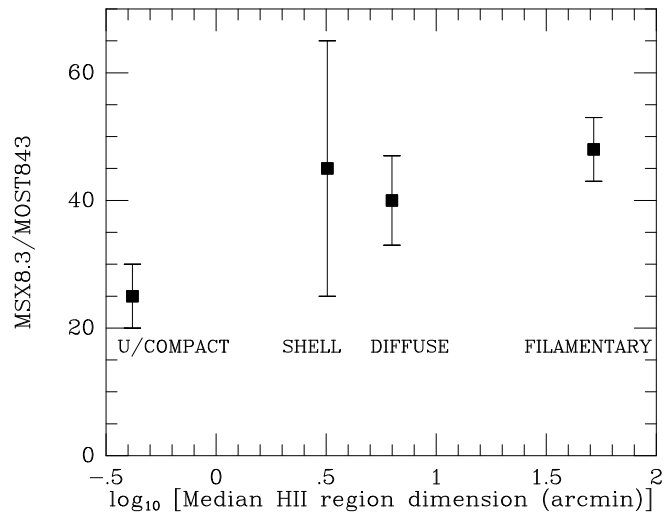}
\caption{As Fig.~\ref{msxmedtyp} but the ratios are plotted against the logarithm of the 
median apparent size of objects in each morphological group.}
\label{msxdim}
\end{figure}

Figures~\ref{msxmedtyp} and \ref{msxdim} seek correlations between MIR/radio flux
densities and the type of H{\sc ii} region, and its characteristic size, respectively.
These figures are derived from the data presented in Table~2.  It is sufficient to
explore any trend with the ratio MSX/MOST, because we know SST and 
MSX fluxes are linearly correlated.  A weak trend toward larger median ratios in
less compact sources is suggested by Fig.~\ref{msxmedtyp} and explored in
Fig.~\ref{msxdim} in which we have attempted to quantify the size of the different
morphology types, using the square root of the median of the areas in each category.
For ultracompact and compact regions we have measured the full width at half maximum
(above the surrounding sky) of the bright cores at 8.0\,$\mu$m and have used this
as the characteristic dimension for these objects.  The median size of these 18 compact
regions is 25$\pm$4\,arcsec (standard error of the median), or 0.42\,arcmin.
The 15 regions whose ratio of MSX/MOST exceeds the median plus 3 standard 
errors of the median (a value of 40) consist of 3 compact, 2 shell, 7 diffuse, and 
2 filamentary sources (excluding the self-absorbed G312.77+0.06).  Each of these 3 compact 
regions also contains filaments, or shell structure or has a MIR halo.  Therefore, this
trend probably reflects the onset of the loss of radio signal with increasing 
apparent size.

It might instead indicate that compact regions other than G312.77+0.06
are dense enough to suffer self-absorption of free-free emission.  Alternatively the
trend might arise because the high density of dust in 
physically small sources leads to stronger MIR thermal continuum emission at the
expense of fluorescent PAH bands.  However, this would also increase the dust emission from
small grains (PAH clusters with hundreds of carbon atoms) observed as a broad plateau 
of emission in the 8.0-$\mu$m {\it IRAC} band, underlying the 6.2, 7.7, and 8.7-$\mu$m bands.
This would offset some of the lost PAH emission.  Detailed modeling of specific regions 
is required to explore this alternative hypothesis, incorporating both
the formation and destruction rates of PAHs and small grains, but this is beyond
the scope of this paper.

There is significant scatter in Figs.~\ref{hiimsx} and \ref{hiisst} but one might have
predicted even more than is observed because variations in excitation parameter and in
843-MHz optical depth from source to source would affect the radio spectral shape and turnover 
frequency.  However, as CG concluded, most of the H{\sc ii} regions in this field are not 
optically thick at 843\,MHz; even the compact regions.  The very fact that MOST detected so 
many regions argues that the old criterion, that below 1\,GHz H{\sc ii} regions are optically
thick, is not sufficient.  The general absence of optically thick regions at 843\,MHz in
the {\it l}=312$^\circ$ field is also borne out by Figs.~\ref{msxmedtyp} and \ref{msxdim}.
The small MIR/radio ratios among compact objects implies there is little loss of 843-MHz
signal.  By comparison with the median of 25 for MSX/MOST for compact regions plotted 
in these figures, the cometary G312.77+0.06 has a ratio of 250, illustrating plausibly
how these ratios would be affected by significant optical depth.

Among the many H{\sc ii} regions in this field is one object that would be defined as
ultracompact, namely G310.69$-$0.31.  The SST 8.0-$\mu$m image has all the characteristics
of a minimally resolved bright source; a size between 1.5$^{\prime\prime}$ and 2$^{\prime\prime}$ (slightly 
in excess of the point spread function's FWHM) and the hexagonal pattern of SST diffraction 
spikes.  The residual image (with the suitably scaled PSF subtracted) shows a central hole at
the core but has a saturated ring immediately around this hole to a radius of roughly 
1$^{\prime\prime}$.  
Its ratio of {\it IRAC}/MSX is 1.16$\pm$0.08, exactly as predicted for a source with the SED of an 
H{\sc ii} region that we have presented in Fig.~\ref{diffcfn}.  This result also confirms that
{\it IRAC} photometry of a region dominated by a pointlike core (although we cannot exclude the 
existence of a very faint halo of emission) is essentially correctly assessed.
Inspection of the Basic Calibrated Data images sugests that the object
is at or near saturation at 8.0\,$\mu$m, but not severely saturated.  Therefore, we believe
there is no major loss of flux in integrating its image and the {\it IRAC}/MSX ratio is credible.
However, the source is too faint to 
have been detected by the LRS instrument.  The ratios of both MSX/MOST (5) and {\it IRAC}/MOST 
(6) are the lowest of any source in Table~1, perhaps because dust competes more effectively 
for stellar UV photons than do PAHs.  

\section{MIR imagery of H{\sc ii} regions}
We present monochrome SST images of twelve objects that span the range of MIR 
morphologies of H{\sc ii} regions in this field.  These are chosen to highlight the
structural MIR characteristics that can be associated with various radio types. 
Although the printed paper presents black and white renditions of the {\it IRAC} images, 
the electronic version shows them as 3-colour pictures in which 4.5-$\mu$m (blue), 5.8-$\mu$m (green),
and 8.0-$\mu$m (red) images have been combined.  Our reasons for this particular trio
are as follows.  Firstly, 3.6\,$\mu$m shows too many stars and clutters the diffuse structure.
MSX was often unable to provide the requisite confirmation of PAHs by the 
test of matching morphologies in its 8.3 and 12.1-um bands (both of which include PAH
band emission) because the sensitivities of all its other bands were so much lower than that
of the 8.3-$\mu$m band.  By contrast, {\it IRAC} offers two sensitive bands capable of sampling 
PAH emission.  In order to confirm the existence of PAHs, that are widely believed to be the
dominant spectral features in diffuse H{\sc ii} regions, we wished to see the same morphology
emerge in {\it IRAC}'s 5.8 and 8.0-$\mu$m bands.  One can also seek subtle changes in PAH 
populations through spatial variations in intensity ratios or, equivalently, in false colour
across regions.  For example, PAH emission from PDRs generally appears yellow or orange 
in our false-colour gallery because both 5.8 and 8.0-$\mu$m bands contain comparably
strong fluorescent PAH emission features.  With {\it IRAC} we still have a probe for H$_2$ lines 
through the 4.5-$\mu$m band too.

Although colour differences across a single H{\sc ii} region have relative significance, 
note that variations between colour images of different regions do not necessarily indicate
any spectral differences between the objects.  Our intent was simply to offer colour images 
that would reveal the intrinsic structure of each region, setting the colour balance and
contrast object by object, rather than aiming at an absolutely black ``sky" for all sources,
or a yellow level for PAHs that is constant for every source.  

\begin{figure}
\vspace{7.5cm}
\includegraphics{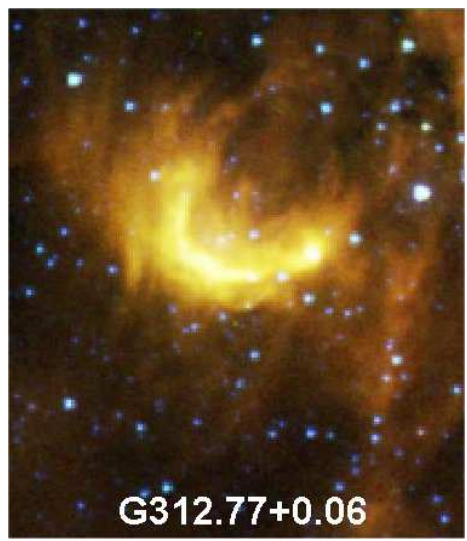} 
\caption{Cometary H{\sc ii} region G312.77+0.06 (MIR morphology type Ch),
with a diffuse halo.  The image
is orientated in Galactic coordinates and measures 205 by 228\,arcsec, {\it l} by {\it b}.}
\label{31277}
\end{figure}

The MIR appearance of the cometary region in our sample  (G312.77+0.06, Ch: 
Fig.~\ref{31277}) clearly resembles the canonical radio structure yet it is also 
associated with a considerable number of filaments.

\begin{figure}
\vspace{7.5cm}
\includegraphics{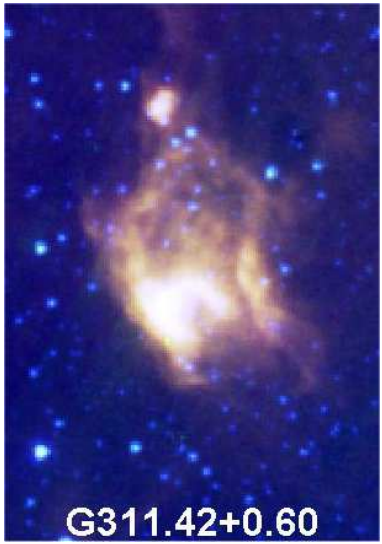}  
\caption{Compact H{\sc ii} region G311.42+0.60 (Cs) with shell and multiple cores.  
Image is 162 by 288\,arcsec.}
\label{31142}
\end{figure}

\begin{figure}
\vspace{8.cm}
\includegraphics{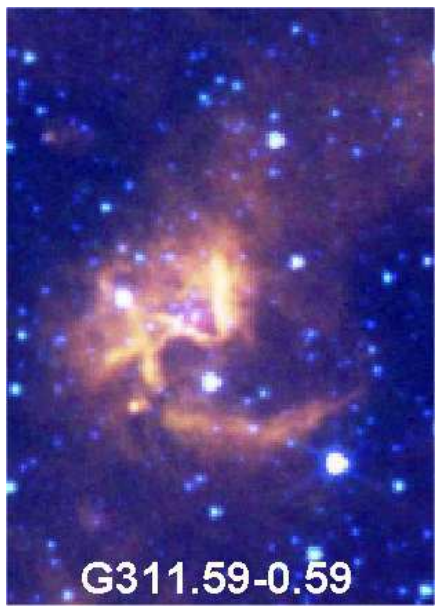}  
\caption{Compact H{\sc ii} region G311.59$-$0.59 (Cf), with its filamentary core. 
Image is 169 by 240\,arcsec.}
\label{31159}
\end{figure}

\begin{figure}
\vspace{8.cm}
\includegraphics{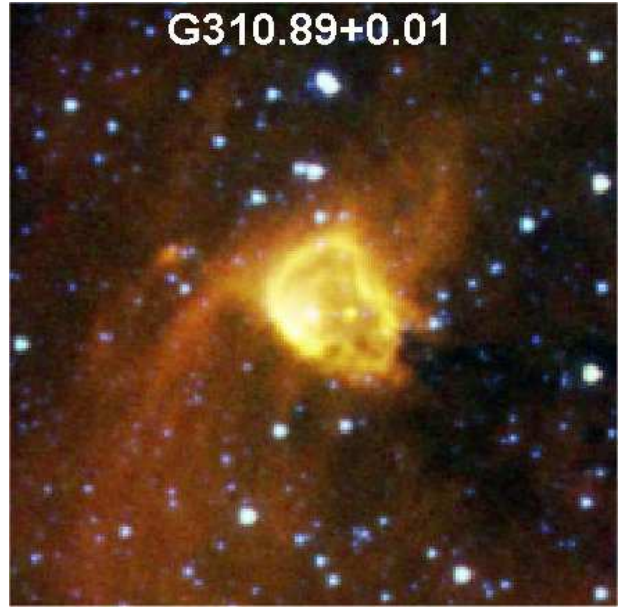} 
\caption{Compact H{\sc ii} region G310.89+0.01 (Cs), with an extended halo in 
the form of a streamer, as well as a bright, shell at its core. Image is 240 by 240\,arcsec.}
\label{31089}
\end{figure}

\begin{figure}
\vspace{7.5cm}
\includegraphics{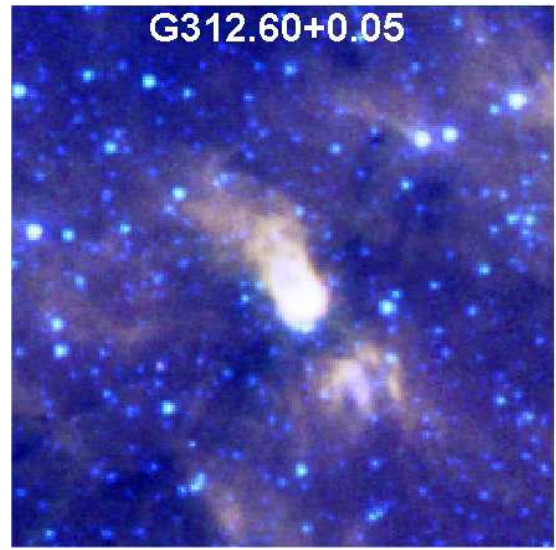} 
\caption{Compact H{\sc ii} region G312.60+0.05 (Ch), with a bright, diffuse halo.  
Image is 240 by 240\,arcsec.}
\label{31260}
\end{figure}

The term ``compact H{\sc ii} region" may be something of a misnomer in the MIR.  Each of our
radio-compact objects does indeed have a MIR-bright core with apparent scale less than 
1$^\prime$.  Some are well separated from the surrounding ISM (G311.42+0.60, Cs: Fig.~\ref{31142}),
but there are often associated multiple centres of emission, intersecting filaments (G311.59$-$0.59, Cf:
Fig.~\ref{31159}), and even extended haloes in which 8.0-$\mu$m emission dominates.
G310.89+0.01 (Cs: Fig.~\ref{31089}) has a halo in the form of an extensive streamer but its
most obvious feature in the MIR is the bright partial shell around its core. 
G312.60+0.06, (Ch: Fig.~\ref{31260}) is a particularly
clear case of a halo around a compact source. It is associated with a small white core (implying
MIR thermal continuum) but it is flanked by quite extensive and relatively bright 
diffuse emission.  The haloes could be emitting in PAH bands, H$_2$ lines or 
simply be due to continuum from warm ($\sim$350\,K) dust grains heated by starlight.  The cores 
of compact H{\sc ii} regions are generally yellow (PAHs) but some are white, indicative of 
the role played by thermal emission from hot dust rather than PAHs or any specific fine structure 
or H$_2$ emission lines.

\begin{figure}
\vspace{8.cm}
\includegraphics{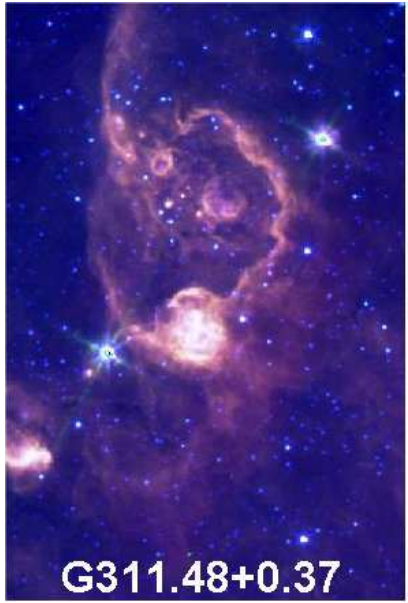} 
\caption{Compact H{\sc ii} region G311.48+0.37 (Cs) which is dominated by
a bright compact region yet incorporates both shells and filaments in the MIR. 
Image is 720 by 480\,arcsec.}
\label{31148} 
\end{figure}

Most shell-type H{\sc ii} regions consist of bounded ionized zones in the radio and in
H$\alpha$ emission.  Their MIR counterparts are filamentary PDRs that enclose the
ionized gas.  Some may include dominant, bright cores (G311.48+0.37, Cs: Fig.~\ref{31148}).  

\begin{figure}
\vspace{8.cm}
\includegraphics{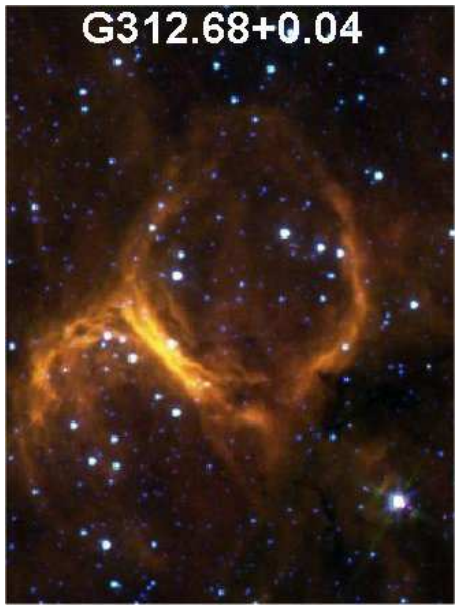} 
\caption{Pair of interacting shell regions with bright interaction bar: 
G312.68+0.04 (Ci) and G312.71+0.02 (Ci). Image is 360 by 480\,arcsec.}
\label{31268}
\end{figure}

\begin{figure}
\vspace{5.5cm}
\includegraphics{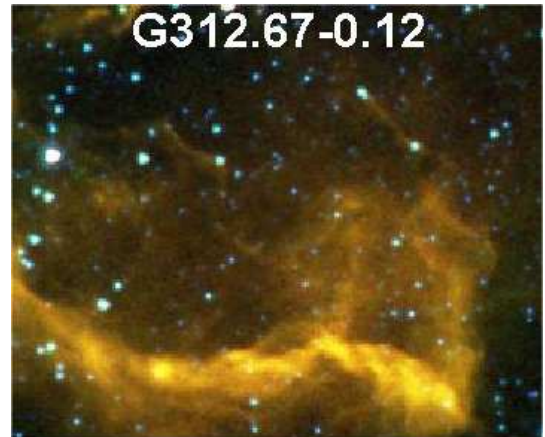} 
\caption{Partial shell source, G312.67$-$0.12 (S). Image is 288 by 240\,arcsec.}
\label{31267}
\end{figure}

\begin{figure}
\vspace{8.0cm}
\includegraphics{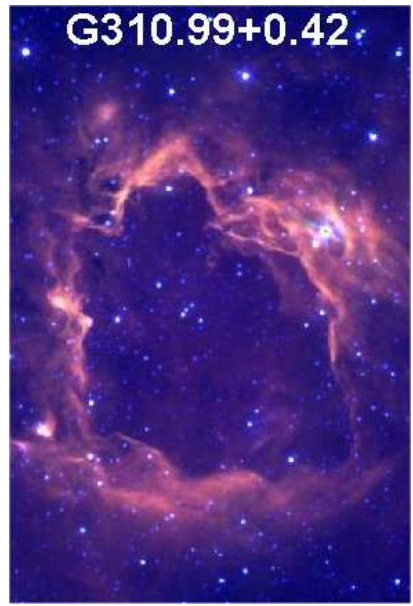} 
\caption{Shell H{\sc ii} region G310.99+0.42 (S).   Image is 480 by 720\,arcsec.}
\label{31099}
\end{figure}

Our sample contains relatively few shell sources.  
G312.68+0.04 (type Si: Fig.~\ref{31268}) is the north-western of a pair of interacting bubbles.  The
second object is G312.71+0.02 (Si), south-east of the bright PDR that arises
from the interaction of the shells blown by the winds from the two exciting stars. The odd structure,
G312.67$-$0.12 (S) (Fig.~\ref{31267}) consists of a bundle of filaments with contiguous diffuse 
emission, suggestive of a partial shell.  
As noted by CG, even at the resolution of MSX, the shell comprising the PDR of G310.99+0.42
(S: Fig.~\ref{31099}) is entirely composed of overlapping filaments.  These are very well shown 
at the higher resolution of SST (see also Fig.~\ref{mira1}).

\begin{figure}
\vspace{7.0cm}
\includegraphics{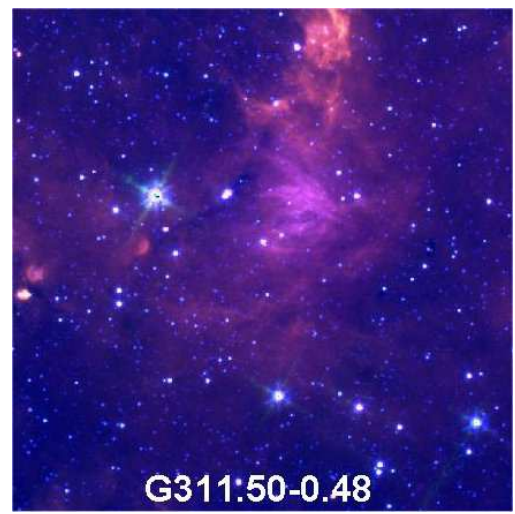} 
\caption{Diffuse H{\sc ii} region G311.50$-$0.48  (D).   Image is 720 by 720\,arcsec.}
\label{31150}
\end{figure}

\begin{figure}
\vspace{5.5cm}
\includegraphics{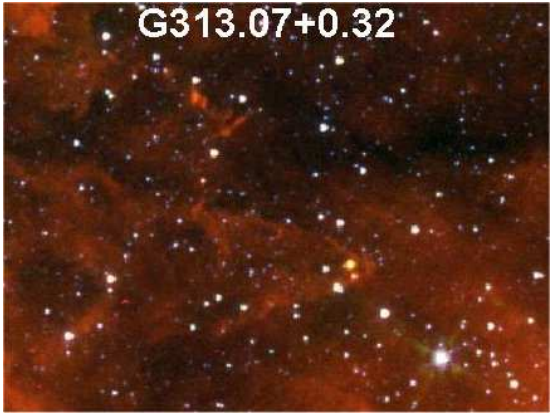} 
\caption{Diffuse H{\sc ii} region, G313.07+0.32 (Df), containing filaments. 
Image is 480 by 360\,arcsec.}
\label{31307}
\end{figure}

Diffuse regions are the second most numerous morphology in the MIR.  Truly diffuse emission
(no filaments, shell, or core) is associated with G311.50$-$0.48 (type D:Fig.~\ref{31150}).  
Most of their MIR emission is diffuse and purple, a colour that might indicate that H$_2$ lines 
dominate both the 4.5 and 8.0-$\mu$m bands.  Its {\it IRAS} LRS spectrum (Fig.~\ref{lrs}) is both 
too noisy and at too poor a spectral resolution to be capable of showing any of these molecular 
lines.  Diffuse regions may contain filaments, such as G313.07+0.32 (Df: Fig.~\ref{31307}). 
Its MIR emission is red because it is detected only at 8.0-$\mu$m.  Fig.~\ref{31307} is a fine
example of an object with so much diffuse PAH emission that it would be almost impossible to
determine where the real sky background occurs within this box, in order to set it to a black level.

\begin{figure}
\vspace{7.0cm}
\includegraphics{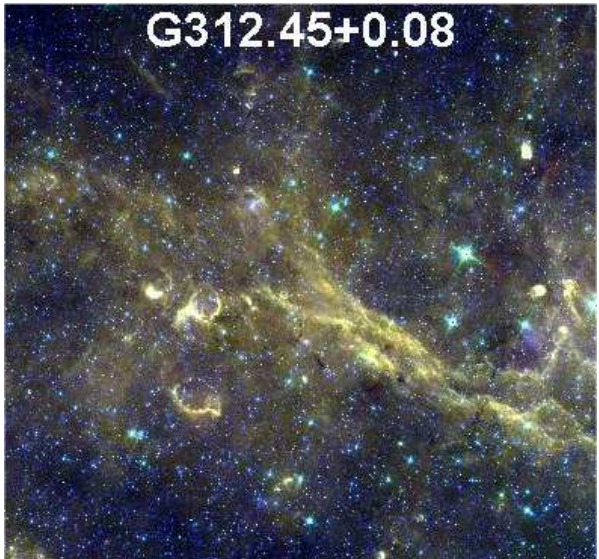} 
\caption{Large filamentary region G312.45+0.08 (F). Image is 70 by 60\,arcmin.}
\label{31245} 
\end{figure}

G312.45+0.08 (Fig.~\ref{31245}) is a spectacular network of filaments over 1$^\circ$
in length, containing a pair of braided filaments as if two very large bubbles have
come into contact without disruption.  The interaction is apparently not vigorous
but has led to a warping of the interface between them (CG).

H{\sc ii} regions with compact MIR morphology all have compact (or ultracompact)
radio morphology.  Filamentary MIR regions are likewise filamentary radio regions. 
But shells and diffuse regions show no unique relationship between radio and MIR morphologies, 
most likely because the resolution of the SST is so much higher than that of the MOST,
but also because PAHs arise in the neutral zones that surround H{\sc ii} regions, giving
them complementary, rather than identical, structures.  

SST's high resolution implies 
that one could characterize H{\sc ii} regions solely using their MIR structure.  
Kurtz et al. (1999) demonstrated that, in the sample of 15 H{\sc ii} regions they studied
in the radio continuum, extended emission is found in 12 objects.  The dynamic range
required to detect these haloes can be assessed from their maps given that their
contour levels correspond to factors of 2, the lowest levels are given, as are the
rms noise values in their Very Large Array (VLA) maps.  Typical dynamic ranges are between 
60 and 500, calculated as the ratio between the compact peak level and the 3-4 rms contour.
The integrated extended emission in a 5\,arcmin square centred on the compact
components ranges from 1 to 15 times that of the peak intensity, in those objects for
which Kurtz et al. (1999) regard a physical association between halo and compact sources
to be possible.  One third (six regions) of our MIR compact sources show extended emission.  
The dynamic range in surface brightness from each 8.0-$\mu$m peak to the faintest extended 
emission clearly detected with {\it IRAC} (all quantities are measured above the sky emission) ranges
from 8 to 50.  The integrated 8.0-$\mu$m brightness of these haloes, compared with a point 
source matched to the compact peaks' surface brightness levels, lies between 10 and 140.  
Thus, the MIR dynamic range probed by the GLIMPSE survey is about 10 times smaller than
that at 3.6\,cm (the wavelength used by Kurtz and his colleagues).  If associated
with the compact regions, the MIR haloes would contribute 10 times more flux relative to their 
compact peaks than at 3.6\,cm.    If the MIR haloes radiated 
thermally then the expected 843-MHz continuum from these haloes would be 36 times smaller 
than that seen by SST (using our median ratio of {\it IRAC}/MOST without recalibration).  The 
3.6\,cm radio emission would be 47 times smaller than at 8.0\,$\mu$m, with the frequency 
dependence of the Gaunt factor for a typical H{\sc ii} region.    
The average rms noise in the VLA survey fields (Kurtz
et al. 1999, their Table~2) is 0.5\,mJy\,beam$^{-1}$ so the 3 rms level is about 
1.5\,mJy\,beam$^{-1}$, equivalent to 38\,MJy\,sr$^{-1}$ at 8.0\,$\mu$m.  The average of
the 8.0-$\mu$m sky-subtracted halo brightnesses for the six regions with MIR type ``Ch" 
is 35$\pm$10\,MJy\,sr$^{-1}$.  Therefore, GLIMPSE at 8.0\,$\mu$m 
provides comparable sensitivity to extended MIR emission around compact H{\sc ii}
regions to that afforded by VLA surveys such as that by Kurtz and colleagues.

Are there MIR haloes akin to the extended weakly ionized radio haloes?   
Within the ionized gas of an H{\sc ii} region PAHs suffer photodestruction by UV photons.  
If any survived, the wind of the central star would rapidly sweep them away into the PDR.
Allain, Leach \& Sedlmayr (1996a) have shown that small PAHs (those with fewer than 50 
carbon atoms) are destroyed by photodissociation on a timescale of a few years in those 
places in which the characteristic MIR bands are seen, specifically for H{\sc ii} regions.  
Cations and partially hydrogenated PAHs may survive for only about a year (Allain, Leach \& 
Sedlmayr 1996b) because they are less stable than the parent molecules.  Consequently,
small PAHs must be made in abundance in locations in which the MIR bands are emitted, for
example in shocks with speeds between 50 and 200 km\,s$^{-1}$ where shattering and
fragmentation create PAH molecules, PAH clusters, and small grains (Jones, Tielens \& 
Hollenbach 1996).
PAHs with more than 50 carbon atoms may survive the UV radiation field 
and even small PAHs might last as long as 1000 years in the diffuse ISM.  Therefore, PAHs
might exist for significant periods in extended haloes that were only weakly ionized and 
far from the dominant source of destructive UV photons.  If all the 
MIR halo emission were due to PAHs then it would be plausible to associate weak diffuse PAH 
emission with extended radio haloes in which low fluxes of UV photons leaking from 
dense ionized cores excite rather than destroy PAHs.  MIR spectroscopy of such
haloes would be definitive. 

\section{G311.70+0.31: a unique structure}
\begin{figure}
\vspace{6.5cm}
\includegraphics{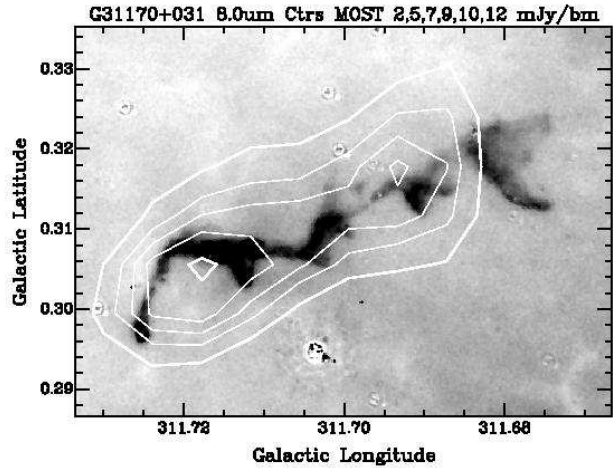}
\caption{Upper: G311.70+0.31 {\it IRAC}\,8.0-$\mu$m residual image. Circular artefacts are
the remnants of point sources that have been subtracted from the original image, in particular
a very bright object at (311.704,0.294).  Lower: G311.70+0.31 8.0-$\mu$m 
image overlaid by 843-MHz contours at levels of 2, 5, 7, 10, 12~mJy~beam$^{-1}$.}
\label{dragon}
\end{figure}

Fig.~\ref{dragon} presents the point-source-subtracted image of a most unusual structure
in this field, weakly detected by MSX, but clearly present in all four {\it IRAC} bands.  Its
spine consists of a series of segments that strongly suggests that a Rayleigh-Taylor 
instability might be responsible.  If so then the mean wavelength of the segments is 
38$\pm$3$^{\prime\prime}$ (standard error of the mean).  However, lacking any estimate
of distance and optical forbidden-line spectroscopy, we are unable to use this information 
on what is surely the fastest growing Rayleigh-Taylor mode to derive a consistent set of
physical parameters for the nebula.  843-MHz radio emission was detected by the MOST and is displayed 
by white contours over the SST 8.0-$\mu$m image.  The object's ratio of MSX\,8.3-$\mu$m
integrated flux to that from the MOST at 843~MHz is 27$\pm$2, entirely consistent with thermal radio 
emission and PDRs associated with PAHs.  The presence of PAHs is confirmed by the identical morphologies
in the {\it IRAC} 5.8 and 8.0-$\mu$m bands.  The radio emission has two peaks, one within the ``bay"
of MIR emission to the east, and a second comparable peak west of the centre.  A physical association 
with the east peak is morphologically stronger.  If the source represented a structure blown 
and ionized by a stellar wind then
one would expect the hot star to betray its presence by a radio continuum peak and by MIR or far-infrared
emission.  A search of the region indicates only a single plausible external source that might be a
star with a wind, namely {\it IRAS}14028$-$6103, a 25-$\mu$m-only object.  Nothing definitive can be
gleaned from the {\it IRAS} photometry or its Low Resolution Spectrum.  No counterparts appear in either
the 2MASS images or those of the Southern H$\alpha$ Survey (Parker et al. 2005), suggesting high 
extinction toward the putative MIR and radio source.

\begin{figure}
\vspace{7cm}
\includegraphics{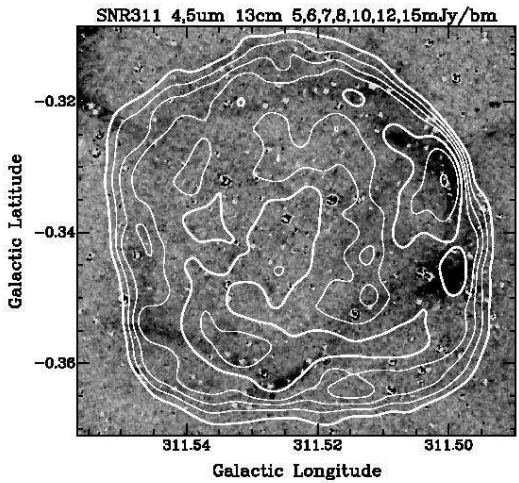}
\caption{SNR G311.52-0.34 4.5-$\mu$m residual image in greyscale, overlaid by 13-cm contours.}
\label{snr2}
\end{figure}

\begin{figure}
\vspace{7cm}
\includegraphics{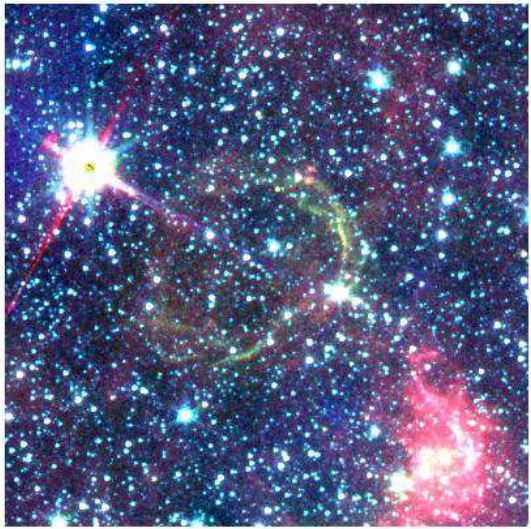}
\caption{SNR G311.52-0.34 false colour image: blue 3.6, green 4.5, red 5.8\,$\mu$m.  The SNR
corresponds to the complete ring around the centre.}
\label{snr3}
\end{figure}

\section{Galaxies, SNRs, and other nonthermal emitters}
GLIMPSE at 8.0\,$\mu$m offers a survey three orders of magnitude deeper than MSX at 8.3\,$\mu$m.
This enables us to probe the MIR/radio ratios for sources other than H{\sc ii} regions.
All five of the suspected background galaxies in the {\it l}=312$^\circ$ field, described by CG, 
were detected by the SST at 8.0\,$\mu$m.  All appear as point sources and so
require no aperture correction.
The ratios of {\it IRAC}/MOST for G310.86+0.01, G311.54$-$0.02, G312.11$-$0.20,
G313.25+0.32, and G313.42+0.09 are 0.080, 0.009, 0.081, 0.019, and 0.034, respectively.
The average ratio for these galaxies is 0.04$\pm$0.01, 900 times smaller than the
empirical ratio for thermal emission. 

If we interpret the extended 8-$\mu$m structure that is associated with radio thermal 
emission as due to PAHs then it is interesting that even diffuse radio regions in 
this field have 8-$\mu$m emission.  This implies that these patchy zones are not 
optically thin to ionizing radiation because they apparently have PDRs.  Seemingly the only
environment in this field that shows a deficit of PAHs (as traced by 8-$\mu$m imagery) is
toward SNRs.

Very few SNRs were detected by MSX (Cohen et al. (2005) present an MSX image of Cas-A) and
nonthermal emission regions are very faint below 9\,$\mu$m, compared with thermal zones.  When
one might suspect that a nonthermal region was detected it is invariably the case
that MSX observed a mixture of thermal and nonthermal filaments but only thermal structures
were detected.  Two (G310.65$-$0.29, G310.80$-$0.41) of the four SNRs in this field are so 
confused by diffuse and filamentary thermal emission that no useful measurements or 
even upper limits can be made of their purely nonthermal portions.  The SNR G312.44$-$0.36
is very large (40\,arcmin) and its interior includes a considerable amount of diffuse emission,
probably thermal, and a considerable number of point and compact sources.  These circumstances
render it a poor candidate for determining the ratio of nonthermal MIR/radio emission.
 
However, the radio-bright young remnant G311.52$-$0.34, discovered by 
Green (1974), is detected by GLIMPSE.  
Fig.~\ref{snr2} overlays the 4.5-$\mu$m residual {\it IRAC} image by 13-cm continuum 
contours (ATCA data kindly provided by Josi Gelfand). 
The spatially integrated, sky-subtracted 8.0-$\mu$m flux density of this SNR can be compared 
with the integrated radio continuum.  The empirical 8.0-$\mu$m to 843-MHz ratio is 0.19, or 0.14 
after aperture correction. WCL list one other nonthermal object, G313.45$-$0.13, but this too lies 
amidst bright thermal filaments.  Combining the ratios for the five galaxies with the aperture-corrected 
ratio for SNR G311.52$-$0.34 yields an average
IRAC/MOST ratio of 0.06$\pm$0.02, over 600 times smaller than the thermal ratio.  On this basis,
the MIR/radio ratio is a sharp discriminant between thermal and nonthermal emission.  

The electronic version
of our paper carries a 3-band false colour image (Fig.~\ref{snr3})
in which 3.6, 4.5, 5.8-$\mu$m emission are
coded blue, green and red, respectively and the remnant can be traced around its complete shell.  
We have sampled the four brightest areas of MIR emission (roughly one in each quandrant
around the elliptical ring) to derive four estimates of {\it IRAC} colours for the SNR. The mean values 
that we calculate yield observed ratios of integrated flux densities at 3.6 and 
5.8\,$\mu$m of 0.17$\pm$0.06, and at 4.5 and 8.0\,$\mu$m of 0.43$\pm$0.06.  After
applying the SSC aperture corrections these become 0.21$\pm$0.07 and 0.55$\pm$0.08,
respectively.  Reach et al. (2006) suggest that shocked gas contributes the {\it IRAC} emission 
from this SNR.  Our colours, within their uncertainties, overlap the predicted
colour-colour box presented by Reach et al. (2006: their Fig.~2) for ionized gas that has suffered 
a fast shock.   The MIR emission is most obvious at 4.5\,$\mu$m,
and somewhat less bright at 5.8\,$\mu$m, due dominantly to Br-$\alpha$ and [Fe{\sc ii}] lines,
respectively, according to Fig.~1 of Reach et al. (2006).  

\section{Conclusions}
We present evidence that the absolute diffuse calibration of SST at 8.0\,$\mu$m is
too high by about 36\,percent.  All measured {\it IRAC} surface brightnesses 
of any diffuse emission region should be scaled down by a multiplicative factor of 
0.74$\pm$0.07, based on the absolutely calibrated MSX images.  This provides a totally
independent confirmation of the separate analyses of {\it IRAC}'s diffuse calibration that are
based upon the fundamental calibration of the instrument (0.737: 
Reach et al. 2005) and detailed modeling of the light distributions of elliptical galaxies 
(0.74$\pm$0.07: Jarrett 2006).  Therefore, the recommendation of the SSC that one should
scale down 8.0-$\mu$m diffuse BCD flux densities by a multiplicative factor of 0.74 
should indeed produce the best absolute estimate of any extended 8.0-$\mu$m emission.

We find that MIR and radio fluxes are linearly related from our regression analysis.
These correlations reflect the intimate association between ionized zones and their 
surrounding PDRs in which polycyclic aromatic hydrocarbon band emission arises.
Our median analysis for H{\sc ii} regions of F$_{8.3\mu m}$/S$_{843~MHz}$ yields 25$\pm$5
(standard error of the median), confirming CG's original derivation.  This ratio of MIR/radio integrated
fluxes is independent of morphology of the ionized region but is valid only for radio
structures smaller than $\sim$24\,arcmin. Aperture synthesis by MOST implies that the telescope
is progressively less sensitive to structure on scale sizes $>$15\,arcmin, diminishing the 
estimated fluxes of larger sources. Compact H{\sc ii} regions below $\sim$3 arcmin in 
size tend to have somewhat smaller MIR/radio ratios, which might be attributed to 
absorption by dust grains of UV photons in some of these small, dense regions.  The
only cometary region in our sample appears to suffer substantial thermal 
self-absorption of its radio flux. 

We find an equivalent correlation for H{\sc ii} regions between observed 8.0\,$\mu$m and 
843-MHz fluxes, albeit with more scatter than for MSX.  With the SST pipeline and calibration 
that were applied to the GLIMPSE images released to the community, we find an
observed value of F$_{8.0\mu m}$/S$_{843~MHz}$~=~36$\pm$13 best represents this 
relationship.  Removing the 36\,percent absolute calibration problem this
corresponds to an aperture-corrected ratio for {\it IRAC}\,8.0/MOST of 27$\pm$10,
very similar to the MSX/MOST criterion for radio thermal emission regions.
Supernova remnants 
studied with GLIMPSE images (Reach et al. 2006) show that none has appreciable MIR synchrotron 
radiation and the dominant post-shock cooling is usually through H$_2$ lines.  This leads to 
much smaller MIR intensities compared with the radio continuum than for thermal sources, 
which are intimately associated with PDRs and their bright and broadband PAH emission features.  The 
MIR/radio flux ratio discriminates between thermal and nonthermal regions.  For the latter, 
the ratio is 0.06$\pm$0.02, some 600 times smaller than for H{\sc ii} regions.
 
The MIR morphology of H{\sc ii} regions does not match their radio structure one-to-one.  This
is most readily seen by examination of Table~1 where the new homogeneous radio and 
MIR morphologies appear side by side.
A diversity of MIR characteristics can be associated with individual radio types.  There is clear
evidence for extended MIR emission surrounding even compact and cometary H{\sc ii} regions.  
Either filaments or diffuse emission may be found around the cores of compact regions, though
some appear truly isolated on the sub-arcminute scale.  When diffuse MIR structure is found
outside the cores of compact H{\sc ii} regions it has the colour of PAH emission.  Consequently,
compact H{\sc ii} regions may indeed have MIR counterparts to the postulated weakly ionized 
radio haloes.  A valid question after examination of these GLIMPSE images is whether there 
are any compact H{\sc ii} regions in the MIR without shells, filaments, and haloes.  
Given the high-resolution of the SST with {\it IRAC} one might wish to develop new quantitative 
criteria for H{\sc ii} regions, in addition to our MIR morphological characterization.

The SEDs of all the H{\sc ii} regions in our sample show little variation in the wavelength range
seen by the four {\it IRAC} bands.  Well-defined colours have been determined from this sample,
and diagnostic boxes proposed for two {\it IRAC} colour-colour planes.  There is good separation in
these plots between the colours of diffuse thermal emission and those of the plethora of MIR 
point sources encountered in the Galactic Plane.  The colours of known pointlike or 
ultracompact H{\sc ii} regions are also sharply distinguished from those of diffuse thermal
regions in these plots.

The advantages of combining MIR and radio aspects of particular sources are illustrated by
the unique MIR structure of G311.70+0.31 in this field.   
{\it IRAC}'s two longest bands offer a highly sensitive tool for
testing for PAH emission by comparing morphologies at different wavelengths that are each dominated
by these molecular bands.  False colour imagery of G311.70+0.31 shows the excellent match 
between the MIR configurations of the source at 5.8 and 8.0\,$\mu$m, strongly suggestive of
the dominance of PAH spectral features in these two images.  The flux ratio, MSX/MOST, is
exactly what one expects for the combination of thermal radio continuum and 8.3-$\mu$m 
emission from PAH features.  CG emphasized the value
of MSX's filters in which PAH structure detected at 8.3\,$\mu$m would be mimicked in the 12.1-$\mu$m
image.  MSX had a large disparity in responsivity between the 8.3-$\mu$m band and all others,
whereas {\it IRAC} offers four sensitive detectors so false colour imagery can achieve its true potential.

\section{Acknowledgments}
MC thanks NASA for supporting this work under LTSA grant NAG5-7936 with UC Berkeley.  We are grateful to 
Tom Jarrett for sharing his extended source calibration results with us, to
Masahiro Tanaka for sending us the complete IRTS spectrum 
of the diffuse ISM, and to Josi Gelfand for sending us the ATCA images of SNR G311.52$-$0.34. 
We thank the anonymous referee for many valuable and detailed comments and suggestions.
This research made use of data products from the Midcourse Space 
eXperiment.  Processing of the data was funded by the Ballistic 
Missile Defense Organization with additional support from NASA Office of Space Science.  
This research has also made use of the NASA/IPAC Infrared Science Archive, which is 
operated by the Jet Propulsion Laboratory, California Institute of Technology, 
under contract with the National Aeronautics and Space Administration.
The MOST is owned and operated by the University of Sydney, with support from the 
Australian Research Council and Science Foundation within the School of Physics.
This research made use of Montage, funded by the
National Aeronautics and Space Administration's Earth Science Technology
Office, Computational Technnologies Project, under Cooperative Agreement
Number NCC5-626 between NASA and the California Institute of Technology.
Support for this work, part of the {\it Spitzer} Space Telescope Legacy Science 
Program, was provided by NASA through contracts 1224653 (Univ. Wisconsin Madison), 
1224681 (Univ. Maryland), 1224988 (Space Science Institute), 1259516 (UC Berkeley), 
with the Jet Propulsion Laboratory, California Institute of Technology under NASA contract 1407. 
R.I. acknowledges funding as a Spitzer Fellow.

\end{document}